\documentclass[showkeys,preprintnumbers]{revtex4}
\pdfoutput=1

\usepackage{graphicx}
\usepackage{amssymb}
\usepackage{amsmath}
\usepackage{multirow}
\usepackage[utf8]{inputenc}
\usepackage[english]{babel}
\usepackage{amsthm}
\newtheorem{pp}{Proposition}[section]
\newtheorem{lm}[pp]{Lemma}
\newtheorem{cl}[pp]{Corollary}
\newtheorem{df}[pp]{Definition}
\begin{document}
\newcommand{\dfn}[1]{\begin{df}#1\end{df}}
\newcommand{\prp}[1]{\begin{pp}#1\end{pp}}
\newcommand{\lem}[1]{\begin{lm}#1\end{lm}}
\newcommand{\crl}[1]{\begin{cl}#1\end{cl}}
\newcommand{\prf}[1]{\begin{proof}#1\end{proof}}
\newcommand{\q}[1]{``#1''}
\newcommand{\lrbr}[1]{\left\lbrace#1\right\rbrace}
\newcommand{\lrsqp}[1]{\left[ #1\right]}
\newcommand{\lrp}[1]{\left( #1\right)}
\newcommand{\lsmatr}[1]{\left\lbrace\begin{smallmatrix}#1\end{smallmatrix}\right.}
\newcommand{\lragl}[1]{\left\langle #1\right\rangle}
\title{A General Theory of Concept Lattice (II):\\ 
\small{Tractable Lattice Construction 
and Implication Extraction}}
\author{Tsong-Ming Liaw}
\email{ltming@gate.sinica.edu.tw} 
\affiliation{Institute of Physics, Academia Sinica, Taipei, Taiwan 11529}
\author{Simon C. Lin}
\email{Simon.Lin@twgrid.org, Simon.Lin@cern.ch}
%\affiliation{Institute of Physics and ASGC, Academia Sinica, Taipei, Taiwan 11529}
\affiliation{Institute of Physics and ASGC, Academia Sinica,  Taipei, Taiwan 11529}
\begin{abstract}
As the second part of the treatise \q{A General Theory of Concept Lattice}, 
this paper speaks of the tractability of
the general concept lattice 
for both its {\it lattice structure} and {\it logic content}.
%The generic concept lattice is a theory more general than the unification of 
%the traditional formal-concept lattice and rough-set lattice.
%This is because one can recover the nodes corresponding to these lattices 
%in the generic concept lattice and, moreover, the generic concept lattice also 
%discloses additional features, from the formal context, which are unobserved by means of 
%formal-concept lattice and rough-set lattice. 
%The generic concept lattice
%is a thorough categorization of object class based on the generalised attributes in $M^\ast$, 
%which are whatever 
%possibly 
%constructed out of the attribute-set $M$, by means of Boolean operations, 
%considered in formal context.  
The general concept lattice permits 
a feasible construction that 
can be completed in a single scan of the formal context, 
though the conventional formal-concept lattice and rough-set lattice 
can be regained from the general concept lattice.
%approaches based on the generic concept lattice 
%become almost trivially tractable.
%is also  in practice, either 
%for finding object-class through attribute-matching 
%or for determining intents through object-matching.
The logic implication deducible from 
the general concept lattice 
takes the form of $\mu_1\rightarrow \mu_2$
where $\mu_1,\mu_2\in M^\ast$ are composite attributes
out of the concerned formal attributes $M$.
Remarkable is that with a single formula based on the 
contextual truth $1_\eta$ 
one can deduce all the implication relations 
extractable from the formal context.\\
For concreteness, it can be shown that any implication $A\rightarrow B$ 
($A,B$ being subsets of the formal attributes $M$)
discussed in the formal-concept lattice 
corresponds to a special case of $\mu_1\rightarrow \mu_2$ 
by means of $\mu_1=\prod A$ and $\mu_2=\prod B$.
Thus, 
%there is no more need of the complicateness 
one may elude the intractability
due to 
searching for the Guigues-Duquenne basis appropriate for 
the implication relations deducible from the formal-concept lattice.
Likewise, one may identify 
those $\mu_1\rightarrow \mu_2$
where $\mu_1=\sum A$ and $\mu_2=\sum B$ with 
the implications that can be acquired from the rough-set lattice.
(Here, the product $\prod$ stands for the conjunction
and the summation $\sum$ the disjunction.)
%deduced from the generic concept lattice
%can be implemented by one single formula but 
%onetheless, more general than the rules deduced 
%from the traditional concept lattices.
%This is in contrast with the study of rules of 
%implication on the traditional formal-concept lattice. 
\end{abstract} 
\keywords{General Concept Lattice;   
Contextual Truth; Pseudo Intent; 
Guigues-Duquenne Basis.}
%\date{May 31, 2019} 
\maketitle
\section{introduction}\label{one}
%The importance of the formal concept analysis ({\bf FCA}) \cite{Wi82,GW99} and the rough set theory ({\bf RST}) \cite{Pa82,Pa91} is nowadays well-recognized, while it is generally believed that {\bf FCA} and {\bf RST} are mutually expressible \cite{Ke96,DGO01,GD02,DG03,YY04,Wa05}.
The {general concept lattice} ({\bf GCL}) \cite{LLJD12-1} 
is a novel structure 
that brings together the {\it formal-concept lattice} (FCL) 
\cite{Wi82,GW99,Wi05} and the {\it rough-set lattice} (RSL) 
\cite{Ke96,GD02,DG03,YY04} on a common theoretical foundation.
The {\bf GCL} accomplishes 
the categorisation of whatever discernible object sets 
into {\it general extent}s 
according to the {\it general intent}s,
which are whatever features the {general extent}s
can possess.
The general intent can be represented by the  
pair of the generalised formal-concept property ({\it Gfcp}) and the generalised rough-set property ({\it Grsp}), 
which are respectively the generalisations for the intents  
of FCL and of RSL \cite{LLJD12-1}.
%The {\bf GCL} furnishs the full categorisation to which   
%both the FCL and RSL  pertain. 
%For the {lattice content}, the {\bf GCL} 
%furnishes the full categorization based on $F(G,M)$, 
%where one can not only recover all the nodes of conventional FCL and RSL but also disclose additonal nodes.
There are well known challenges
both for the construction of lattice 
\cite{Ku01,KO02}
and for extracting the logic content in FCL
\cite{GW99,GD86,Kso04,KO08,Sb09,Df10,DS11,BK13}.
In this paper, one will show that 
the {\bf GCL} is free from such intractability 
problems, 
%either in constructing the lattice or in extracting its 
%logic content, 
the {{\bf GCL}} 
is in practice much easier 
to handle than the FCL and RSL.
To begin with, a remarkable principle 
that leads to the {\bf GCL} is that 
the information content revealed by the formal context $F(G,M)$
should not be specific to the  
particular choice of formal attribute set $M$.
%This intuitively agrees with one's common sense.
%For example, telling \q{the object $1$ is red ($R$), the object $2$ is green ($G$) and the object $3$ is blue ($B$)} 
%should be not different from 
%telling \q{the objects $1,2$ are either red or green ($R+G$), the objects $2,3$ are either green or blue ($G+B$) and 
%the objects $1,3$ are either red or blue ($R+B$)}.
%Moreover, 
%Since \q{being composite or non-composite attribute}
%should not be fundamental to the property of objects, 
In effect, the possible property characterisations for the objects
may run over 
%all the attribute composition 
%out of $M$ by means of Boolean operations, namely,  over 
the generalised attribute set $M^\ast$. 
Hence, by consistency, the {\bf GCL} turns out to %make reference 
depend on 
the {\it extended} formal context {$F^\ast(G,M^\ast)$},
see Lemma~2.8~in Ref.~\cite{LLJD12-1}, 
since $F^\ast(G,M^\ast)$ is the direct consequence of $F(G,M)$. 
Intriguingly, 
despite the enormous increase of attribute freedom, owing to 
the extension from $M$ to $M^\ast$, 
one finds that the problem treatment
is much simplified.  
Here, an interpretation
for the {\bf GCL} to be manageable 
is that $F^\ast(G,M^\ast)$
unveils additional instructive relations  
which are not accessible otherwise.
%if the composite attributes are elided.
Clearly, such instructive relations
can by no means be observed in the FCL and the RSL
that only %ostensibly 
refer to $F(G,M)$. 

In Section~\ref{two}, one will clarify 
%the generality of {\bf GCL}
%in relation to the other lattices, including the 
%FCL and RSL, one may 
%obtain from the formal context.
%It will be shown that  
the technical {\it origin} why {\bf GCL} is tractable
both in constructing the lattice structure and 
in implementing the logic content.
Basically, the {\bf GCL} comprises $2^{n_F}$
nodes (general formal concepts) characterising 
all the distinct object classes (general extents) 
discernible by the formal context, see Proposition~3.4 and 3.5 of Ref.~\cite{LLJD12-1}, thus, 
the efforts in deciding which object class
is categorised are in fact inessential. 
Further, it will also become obvious that 
the identifications for the %$2^{n_F}$ 
general intents
in fact exhaust all the $|M^\ast|$ generalised attributes. 
In effect, corresponding to each general extent {$X\in E_F$}, 
see Proposition~3.5 and Proposition~3.6 of Ref.~\cite{LLJD12-1}, 
the general intent amounts to the closed interval
$[X]_F=\lbrace \mu\in M^\ast \mid\ \mu^R=X \rbrace=[\eta(X),\rho(X)]$, 
which has the upper bound {$\rho(X)$} (%the generalised rough-set property 
{\it Grsp}) and the lower bound {$\eta(X)$} (%the generalised formal-concept property 
{\it Gfcp}).
Note that the whole generalised attribute set $M^\ast$ 
is then divided into $2^{n_F}$ non-overlapping 
general intents on the {\bf GCL}, 
in contrast to the cases for FCL and RSL 
where attributes in $M$ can be {\it repeatedly} used for the intents. 
Therefore, it is a
%fair to anticipate the 
simplicity 
to determine the general intents 
since each of them appears to be a unique closed 
interval up to its bounds,
which are ordered according to Galois connection. 
Moreover, one notices that 
the complexity in determining all the bounds
can still be halved
%as  
%only either of their bounds are in fact needed
by virtue of the conjugateness relation of Proposition~3.7 in 
Ref.~\cite{LLJD12-1}.
%($[X_1]_F\cap [X_2]_F=\emptyset$ iff $X_1\neq X_2$ for $X_1, X_2$ being general extents) 
%the {\bf GCL} 
%essentially comprises the generalised versions of {FCL and RSL}.
%While 
%Note that there exist {\it generalised} {FCL and RSL} 
%which are obtained respectively from {FCL and RSL} 
%as the direct consequence of {the extension \q{$F(G,M)$ to $F^\ast(G,M^\ast)$}.Remarkably, while the {\bf GCL} can be regarded as the 
%association of the generalised {FCL and RSL}, unlike both of them, 
%the general intents are not nested in the sense 
%\q{the attributes of $M^\ast$ 
%are used as intents in a multiple manner}.
Remarkable is also that the 
the non-overlapping of general intents
facilitates the implementations of logic implications.
%On one hand,
The idea, relating two attributes pertaining to two general intents 
 whose general extents are ordered via set-inclusion relations by an implication,
%(not only) 
maps to 
%the idea applying to 
the extraction of implications from the FCL \cite{GD86} and RSL.
%(but also onto a similar idea with which one could have extracted implications from the RSL). 
On the other hand,
the logic content of {\bf GCL} are then clarified by means of
the simple equivalence among attributes grouped into the same general intent  
since the equivalent attributes are %simultaneously 
the property of the same object class.

Sec.~\ref{three} is devoted to  
the practical aspect in determining the {\bf GCL}'s lattice structure.
% also in terms of explicit examples. 
Unlike Ref.~\cite{LLJD12-1},
one instead adopts the disjunctive normal form ({\bf DNF}) 
for {\it Gfcp} and the conjunctive normal form ({\bf CNF}) for {\it Grsp}.
%(Definition 2.14, Lemma 3.8, 3.11 \cite{LLJD12-1}).
%It is demonstrated that 
%the new considerations are equivalent 
%to the earlier constructions which are per 
%irreducible attributes.
%1{\bf GCL} is isomorphic to the Boolean algebra formed by a power-set.
It will be demonstrated that
the {\bf GCL} subject to the formal context
is characterised by a simple $\eta$-representation or
$\rho$-representation which is obtained after one single glance over
the formal context.
From the
$\eta$- or $\rho$-representation 
one may read out
the general intent for any given general extent in $E_F$.
Every $n_F$-bit binary string $B_X$ can 
%unambiguously
then be employed to
encode a general formal concept as
$(X,\rho(X),\eta(X))=(X,[B_X]_\rho,[B_X]_\eta)$.
The construction of {\bf GCL} is thus as tractable 
as grouping the objects according to attributes in 
the formal context.
One will also demonstrate  
that both the FCL and RSL can be rediscovered from the {\bf GCL} 
as particular cases:
The FCL categorises 
those object classes which are expressible 
in terms of an {\it intersection} of $m^R$s for some $m$s in $M$ where the whole object collection $G$ is also regarded as an FCL extent, cf. e.g. Ref.~\cite{Wi82};
the RSL categorises those object classes given in terms of a {\it union} of $m^R$s for some $m$s in $M$ where the empty object set $\emptyset$ is also regarded as an RSL extent, cf. e.g. Ref.~\cite{YY04}.

In Sec.~\ref{four}, an implication relation
discussed in the frame of {\bf GCL} 
is considered between two different attributes, say 
$\mu_1\rightarrow \mu_2$, which
can be intuitively 
mapped onto the contextual Venn diagram (Definition~2.9 of Ref.~\cite{LLJD12-1}), giving rise to 
the set inclusion relation $\mu_1^R\subseteq \mu_2^R$. 
%where $\mu_i^R$ denotes the object set 
%which possesses the attribute $\mu_i$, for $i=1,2$. 
Interestingly, 
the relation %the conventional employment of
$A{\rightarrow} B\ 
%\mbox{with}\ 
(A\subseteq M,B\subseteq M)$
%\lsmatr{A\in 2^M\\
%B\in 2^M}$ 
developed from FCL \cite{GD86}, i.e. $A\stackrel{fcl}{\rightarrow} B$, 
appears to be a special case
of the logic implication from {\bf GCL}. 
Likewise, one may also recognise the relation $A\stackrel{rsl}{\rightarrow} B$, which is the one that could have been developed from RSL,
as a special case
of the {\bf GCL}-based rules of implication. 
%$A\stackrel{rsl}{\rightarrow} B \iff \mu_1\rightarrow\mu_2\ \mbox{with}\ \lsmatr{\mu_1=\sum A\\ \mu_2=\sum B}$.
It is noteworthy that 
the logic implication arising from the {\bf GCL}
can be deduced from one single formula
%, i.e., $\forall \mu\in M^\ast\ \mu\rightarrow \mu\cdot 1_\eta$, 
depending on the contextual truth $1_\eta$, see {Definition~2.5} of Ref.~\cite{LLJD12-1}. 
%corresponds to the sum of all the attribute components within the $\eta$-representation. (Also, $1_\eta$ itself may serve as a logic condition which can be employed as a criterion that judges whether any implication rule can exist.) 
Remarkably, one may then forgo 
the prevalent process of finding pseudo-intents for 
the construction of the minimal implication bases, 
the so-called Guigues-Duquenne bases or stem bases \cite{GW99,GD86,Kso04,KO08,Sb09,Df10,DS11,BK13}.
Not to mention that  
there are still implication relations deducible from the {\bf GCL}
which can be neither identified with
$A\stackrel{fcl}{\rightarrow} B$
nor with $A\stackrel{rsl}{\rightarrow} B$.

\section{generality versus tractablity}\label{two}
Following the convention of the previous work, the formal context 
is denoted as $F(G,M)$, where $G$ represents the {\it formal objects} 
and $M$ represents the {\it formal attributes}.
The fundamental operation $R$ defines the map
{\it from an object element to an attribute subset} 
and 
{\it from an attribute element to an object subset}:
\begin{equation}
g\in G \mapsto  g^R \in 2^M\ (m\in M \mapsto  m^R \in 2^G), 
\label{eq1}
\end{equation}
based on which the derivation operators 
{$I,\Box$ and $\Diamond$} are given by Eq.~(4)-(6) and (8) of Ref.~\cite{LLJD12-1}.
Note that the treatment of objects and 
attributes are formally {\it different}.
The objects are distinct 
entities while the 
attributes can overlap per {\it conjunction},
a point of view
giving rise to the formal context 
extension from $F(G,M)$ to $F^\ast(G,M^\ast)$
(Lemma~2.8 of Ref.~\cite{LLJD12-1}).
%unless they are assumed to be all mutually exclusive.
%are given as
%\begin{eqnarray}
%X^I&=&\lbrace m\in M \mid gRm,\ \forall g \in X \rbrace,\ Y^I=\lbrace g\in G \mid gRm,\ \forall  m\in M \rbrace,\nonumber\\ 
%X^{\Box}&=&\lbrace m\in M \mid \forall g \in G, gRm \Rightarrow g\in X \rbrace,\
%Y^{\Box}=\lbrace g\in G \mid \forall m \in M, gRm \Rightarrow m\in Y \rbrace,
%\nonumber\\
%X^{\Diamond}&=& \lbrace m \in M \mid \exists g\in G, (gRm,\ g\in X) \rbrace,\ 
%Y^{\Diamond}= \lbrace g \in G \mid \exists m\in M, (gRm,\ m\in Y) \rbrace.
%\label{eq:derivation_O}
%\end{eqnarray}
%Upon employing what one conventionally considers as objects and attributes
%with $U$ and $V$, respectively,
%In other words, the {\bf GCL} based on $(U,V)\equiv (G,M)$ 
%cannot be the same as the {\bf GCL} corresponding to $(V,U)\equiv (G,M)$. 
%can be identified as
%the so-called {\it object-oriented} ({\it property-oriented}) scheme. 
However, there are then
two distinct formal contexts to be considered subject to the same data structure \cite{YY04}, as are  
%for "$U=\lbrace 1,2,3,4,5,6\rbrace$" and "$V=\lbrace a,b,c,d,e\rbrace$"
exemplified and explained in Table \ref{table:formal_scheme}.
%Notably, the employment 
%\q{{\it formal} objects and {\it formal} attributes}, rather than 
%\q{objects and attributes}, for the {\bf GCL} is to stress
%the different manners of handling $G$ and $M$ \cite{LLJD12-1}.\\
%Here, since  
%the treatments for $G$ and $M$
%are different in {\bf GCL}, 
%it seems that one should discuss the {\bf GCL} in two distinct schemes.
%However, it remains an open question
%whether concrete interpretations or even compromises for both schemes
%can be (or should be) simultaneously found.
%\\
Now that if a set of given parameters 
is regarded as objects it can no more be 
regarded as attributes and vice versa,
it remains an open question
whether or how the both formal contexts 
can co-exist in the same treatment,
in what follows one will concentrate 
on a definite choice of objects and attributes. 

\begin{figure*}   
\includegraphics[scale=0.4,angle=-90]{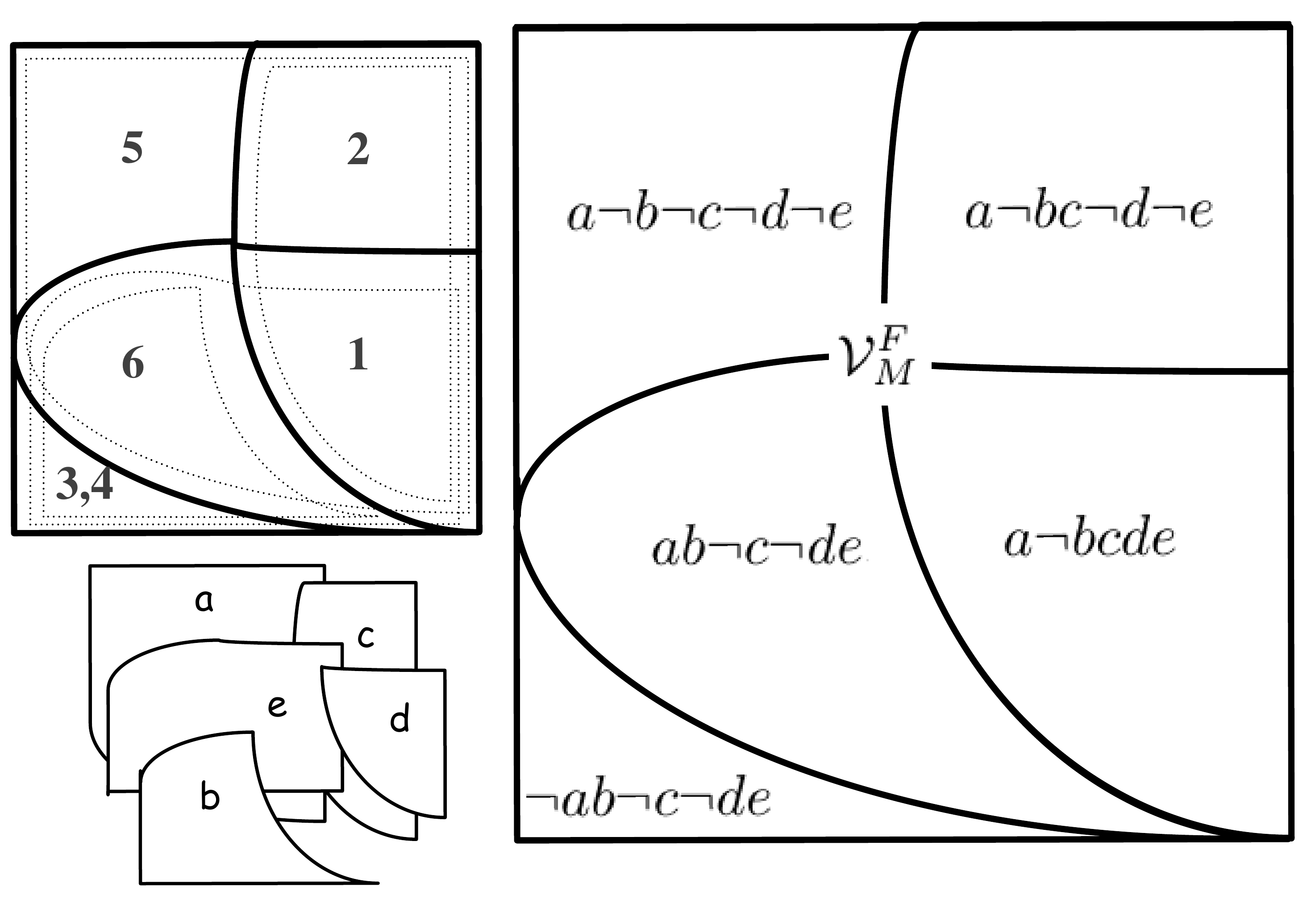}
\caption{Drawing the Contextual Venn diagram 
for the $F(G, M)$ in Table~\ref{table:formal_scheme}\\ 
The full objects of interest 
$G$ are partitioned into $n_F=5$ classes, 
which are discernible by the formal
context, say $\lrbr{D_1,D_2,\ldots D_5}$.
Here, one determines 
the relations among all the attributes  
on ${\cal V}_M^{F}$ by means of $\lbrace m^R\mid m\in M\rbrace$ subject to $F(G,M)$.
%, i.e. by exactly encircling the objects possessing the attribute $m$.
For example, 
\q{$a$} corresponds to the region 
covering $a^R=\lbrace 1,2,5,6\rbrace$ and so forth.
Moreover, for each disjoint region $D_k$ one has  
$\eta(D_k)=\prod_{m\in M} \alpha_m$, where
$\alpha_m =m$ for $m^R\supseteq D_k$ and $\alpha_m=\neg m$ otherwise.
For example, $\eta(\lrbr{1})=a\neg bcde$ and so forth, as is depicted on the rightmost diagram.}
\label{figure:context_venn} 
\end{figure*}

Since the {\bf GCL} makes reference on $F^\ast(G,M^\ast)$,
it is intriguing whether the problem complexity of {\bf GCL} can be reduced 
despite the domain extension {\it from $M$ to $M^\ast$}.
To this end, it is rather instructive to take into account two ordering systems one may establish among attributes.
Without loss of generality, one could assume that
no constraint has been pre-imposed on the given attribute set $M$.
As the first ordering system, one then ends up the conventional Venn diagram ${\cal V}_M^0$, which is 
divided into $2^{|M|}$ disjoint regions in the sense that no pair of attributes in $M$ has an empty {\it intersection}.
Moreover, as the second ordering system, the contextual Venn diagram 
${\cal V}_M^F$ (Definition~2.9 in~Ref.~\cite{LLJD12-1})
is employed to govern 
the attribute relations inferred from the formal context. 
In Fig. \ref{figure:context_venn}, the $n_F$ disjoint regions on ${\cal V}_M^F$ correspond to the $n_F$ object classes 
which are discernible subject to the formal context.
The dimension reduction from $|M^\ast|$ to $2^{n_F}$ 
can thus be relaised by mapping the $|M^\ast|$ generalised attributes
in ${\cal V}_M^0$ 
onto $2^{n_F}$ attribute classes
in ${\cal V}_M^F$, as is
illustrated in Table \ref{table:equivalent_class}.
Notice that each of the $2^{n_F}$ classes is given with respect to 
a definite {\it general extent} $X\in E_F$ as 
$[X]_F=\lbrace \mu\in M^\ast\mid \mu^R=X\rbrace$
(Definition~2.11~in~Ref.~\cite{LLJD12-1}), where
$E_F$ includes all the possible unions of the members in $G_{/R}$ ($|G_{/R}|=n_F$ by {Definition~2.4}~of Ref.~\cite{LLJD12-1}).
%, which
%consists of all the smallest possible subsets of $G$ that can be written in terms of $\mu^R$ with some $\mu\in M^\ast$.
%Therefore, %it is more appropriate to
Adopting $[X]_F$ as the {\it intent of general concept} 
turns out to be more intuitive  
than adopting the pair $(\rho(X),\eta(X))$ 
since $[X]_F$ is itself an equivalent class.
\prp{
%{\bf Proposition~2.1}\\*
Subject to the formal context $F(G,M)$, 
the 2-tuple $(X,[X]_F)$ with $X\in E_F$
is more appropriate than $(X, \rho(X),\eta(X))$ (Proposition~3.4 in~Ref.~\cite{LLJD12-1})
for the general concept framework, where
$[X]_F$ is referred to as the {\it general intent}.
The following statements concerning the general concept framework are in order.
%due to the following facts.
%It can be shown that 
\begin{itemize}
\item
$[X]_F$ can be deduced from $\rho(X)$ and $\eta(X)$ without any additional assumption.\\
%$[X]_F$ can be deduced from $\rho(X)$ and $\eta(X)$ by means of
$[X]_F=[\eta(X),\rho(X)]$, where $[\eta(X),\rho(X)]=\lbrace\mu\in M^\ast\mid\ \eta(X)\leq\mu\leq\rho(X)\rbrace$.
\item
The {\bf GCL} exhausts whatever attributes 
from $M^\ast$,
%one may construct out of $M$.\\ 
$M^\ast=\bigcup_{X\in E_F} [X]_F$. 
\item
Different general intents do not overlap: 
$\forall X_i\forall X_j \in E_F\ [X_i]_F\cap [X_j]_F=\emptyset$ iff $X_i\neq X_j$.
\item
The general concept $(X,[X]_F)$
can be employed as the node on the lattice
since the ordering can be unambiguously defined.
Denoting the nodes $l_i$ and $l_j$ with $\lsmatr{l_i=(X_i,[X_i]_F).\\
l_j=(X_j,[X_j]_F)}$, 
\[l_i\neq l_j \iff \left\lbrace
\begin{smallmatrix}
X_i\neq X_j\\
[X_i]_F\neq [X_j]_F
\end{smallmatrix}\right.,\ l_i<l_j\iff \left\lbrace \begin{smallmatrix}
X_i\subset X_j\\
[X_i]_F<[X_j]_F
\end{smallmatrix}\right.
\]
%$l_i\neq l_j$ iff 
%$\left\lbrace \begin{smallmatrix}
%X_i\neq X_j\\
%[X_i]_F\neq [X_j]_F
%\end{smallmatrix}\right.$ 
%$l_i<l_j$ iff 
%$\left\lbrace \begin{smallmatrix}
%X_i<X_j\\
%[X_i]_F<[X_j]_F
%\end{smallmatrix}\right.$.
\end{itemize}}
%{\bf Proof}:\\ 
\prf{
The closed interval $[\eta(X),\rho(X)]$ is well-defined since $\forall X\in E_F$ $\eta(X)=\prod [X]_F\leq \sum [X]_F=\rho(X)$.
Here, the general intent also includes all the attributes lying between $\eta(X)$ and $\rho(X)$. 
%It is clear that 
%$\forall X\in E_F$, $\eta(X)=\prod [X]_F\leq \sum [X]_F=\rho(X)$, thus,
%the closed interval $[\eta(X),\rho(X)]$ is well-defined. 
%This is to show that $(X,[X]_F)$ is well-defined up to $(X,\rho(X),\eta(X))$ and is thus not ambiguous.
\begin{itemize}
\item
%The statement is equivalently 
%$\mu \in [X]_F$ iff $\eta(X)\leq\mu\leq\rho(X)$. 
%Since 
%$\left\lbrace \begin{smallmatrix}
%\rho(X)=\sum [X]_F\\
%\eta(X)=\prod [X]_F
%\end{smallmatrix}\right.$, 
%This is in effect telling that 
%$[X]_F$ can be deduced from $\rho(X)$ and $\eta(X)$ without any additional assumption.\\ 
$\forall \mu \in [X]_F=\lbrace \mu\in M^\ast\mid \mu^R=X\rbrace\ 
\rho(X)\geq \mu\geq \eta(X)\ 
\therefore\mu \in [X]_F\implies \mu \in [\eta(X),\rho(X)]$.
On the other hand, $\mu \in [\eta(X),\rho(X)]\implies 
\left\lbrace \begin{smallmatrix}
\mu^R\supseteq \eta(X)^R=X\\
\mu^R\subseteq \rho(X)^R=X
\end{smallmatrix}\right.$ (Proposition~3.4~in~Ref.~\cite{LLJD12-1}), 
which implies that $\mu^R=X$, i.e., $\mu \in [X]_F$.
%$\therefore \mu \in [\eta(X),\rho(X)]\implies\mu \in [X]_F$.
Therefore, $\mu \in [X]_F\iff\eta(X)\leq\mu\leq\rho(X)$. %$\blacksquare$
%\item
%Clearly, $[X_i]_F\cap [X_j]_F=\emptyset$ iff $X_i\neq X_j$ for $X_i$ and $X_j\ \in E_F$ by means of the definition $[X]_F=\lbrace %\mu\in M^\ast\mid \mu^R=X\rbrace$.
%Moreover, $\forall \mu \in M^\ast$, $\mu^R=X$ for some $X\in E_F$ since $E_F=\lbrace \mu^R\mid\ \mu\in M^\ast\rbrace$
%(Lemma 3.4 \cite{LLJD12-1}). %$\blacksquare$
\item 
%Basically, $M^\ast\supseteq\bigcup_{X\in E_F} [X]_F$ since $\forall X\in E_F\ [X]_F\subseteq M^\ast$.
Since $\forall \mu\in M^\ast\ \mu^R\subseteq G$ 
({Definition~2.7},~{Lemma~2.8}~in~Ref.~\cite{LLJD12-1})
and $E_F=\lrbr{\mu^R\mid \mu\in M^\ast}$,
$\bigcup_{X\in E_F} [X]_F=\bigcup_{\mu\in M^\ast} [\mu^R]_F=M^\ast$. 
%$\blacksquare$
\item
%For any $X_k \in E_F$ one may let 
% $[X_k]_F$ be represented by 
%$[\mu_k^R]_F$ with
%$\mu_k$ being {\it some} attribute in $[X_k]_F$.
%\end{smallmatrix}\right)$
%be represented by 
%$\left( \begin{smallmatrix}
%[\mu_i^R]_F\\
%[\mu_j^R]_F
%\end{smallmatrix}\right)$, 
%where $\left( \begin{smallmatrix}
%\mu_i\\
%\mu_j
%\end{smallmatrix}\right)$
%is considered as {\it some} attribute in 
%$\left( \begin{smallmatrix}
%[X_i]_F\\
%[X_j]_F
%\end{smallmatrix}\right)$.
%Namely, $\mu_i^R=X_i\ \forall \mu_i\in [X_i]_F$ and $\mu_j^R=X_j\ \forall \mu_j\in [X_j]_F$.
$[X_i]_F\cap [X_j]_F=\emptyset$
contradicts $X_i=X_j$, which implies that 
$[X_i]_F=[X_j]_F$, $\therefore\ [X_i]_F\cap [X_j]_F=\emptyset\implies X_i\neq X_j$. 
% if $[X_i]_F\cap [X_j]_F=\emptyset$, assume by contraries that $X_i=X_j\equiv X$. 
%Then, $\forall \mu_i\in [X_i]_F\ \forall \mu_j\in [X_j]_F$, $\mu_i\neq \mu_j$ but $\mu_i^R=\mu_j^R=X$, which implies $[X]_F\supseteq [X_i]_F\cup [X_j]_F$.
%This is contradictory since, e.g., 
%$[X_i]_F\stackrel{X_i=X}{=}\lbrace \mu\in M^\ast \mid \mu^R=X\rbrace$ 
%is defined to complete all the $\mu$s s.t. $\mu^R=X$, in $M^\ast$, however, it is now told that $[X_i]_F$ can be enlarged.
On the other hand,
if $X_i\neq X_j$ then 
$\mu_i^R\neq \mu_j^R$
%$\forall \mu_i\in [X_i]_F\ \forall \mu_j\in [X_j]_F$ 
and hence
$\mu_i\neq \mu_j$ (cf. Lemma 2.10~in~Ref.~\cite{LLJD12-1}), $\forall \mu_i\in [X_i]_F\ \forall \mu_j\in [X_j]_F$.
Consequently, $X_i\neq X_j\implies [X_i]_F\cap [X_j]_F=\emptyset$.
Therefore, $[X_i]_F\cap [X_j]_F=\emptyset\iff X_i\neq X_j$.
%$\blacksquare$   
\item
%$l:=(X,\rho(X),\eta(X))$ is not ambiguous (Lemma 3.14 \cite{LLJD12-1}).
%On the other hand, consider $l:=(X,[X]_F)$.
$(X_i,[X_i]_F)\neq (X_j,[X_j]_F)\iff X_i\neq X_j$ and $[X_i]_F\neq [X_j]_F$
since $[X_i]_F\cap [X_j]_F=\emptyset\iff X_i\neq X_j$.
%Moreover, whenever $[X_i]_F$ and $[X_j]_F$ are disjoint,
Consider then $[X_i]_F<[X_j]_F$ as 
$\left\lbrace \begin{smallmatrix}
\rho(X_i)<\rho(X_j)\\
\eta(X_i)<\eta(X_j)
\end{smallmatrix}\right.$ since $X_i\subset X_j\iff\left\lbrace \begin{smallmatrix}
\rho(X_i)<\rho(X_j)\\
\eta(X_i)<\eta(X_j)
\end{smallmatrix}\right.$(Proposition~3.14~in~Ref.~\cite{LLJD12-1}).
Therefore, 
$l_i<l_j \iff
\left\lbrace \begin{smallmatrix}
X_i\subset X_j\\
[X_i]_F<[X_j]_F
\end{smallmatrix}\right.$.
%, where the {RHS} is identified as $ 
%\left\lbrace \begin{smallmatrix}
%X_i<X_j\\
%\rho(X_i)<\rho(X_j)\\
%\eta(X_i)<\eta(X_j)
%\end{smallmatrix}\right.$. 
%$\blacksquare$
%Writing in terms of the triplet $(X,\rho(X),\eta(X))$ 
%specialises the bounds of $[X]_F$. 
\end{itemize}
}
While the general extents are all the object classes discernible from the perspective of the formal context, 
every general intent $[X]_F$ collects the attributes 
all the members of the general extent $X$ possess in common,
where $\rho(X)$ and $\eta(X)$ happen to be the upper and lower bound of 
$[X]_F$, respectively.
The general concept $(X,\rho(X),\eta(X))$ 
thus far relates to the RSL-concept and FCL-concept as follows.  
\begin{itemize}
\item
If $X$ appears to be an RSL extent, its corresponding intent 
$X^\Box=\lbrace m \in M\mid\ m^R\subseteq X \rbrace$, see 
Eq.~(8) of Ref. \cite{LLJD12-1}, collects all the {\it unique} attributes 
in $M$ which are not observed on the members of $X^c$.
Since carrying any  
part of these attributes suffices to ensure that an object 
belongs to $X$, the {\it rough-set property} 
as the logical {\bf OR} of the members in $X^\Box$
faithfully characterises the RSL intent. 
Subsequently, the {\it general} rough-set property ({\it Grsp})
can be deduced via the
extension {from $F(G,M)$ to $F^\ast(G,M^\ast)$} as
\begin{equation}
\rho(X)=\sum X^{\Box^\ast}=\sum\lbrace \mu \in M^\ast\mid\ \mu^R\subseteq X\rbrace=\sum_{X_0\subseteq X}\left(\sum [X_0]_F\right)\equiv \sum [X]_F.\label{eq:rhoX}
\end{equation}
\item
If $X$ is an FCL extent $X$,
its intent collects all the attributes in $M$ 
possessed {\it in common} by $X$, say
$X^I=\lbrace m \in M \mid\ m^R\supseteq X \rbrace$. 
Since members in $X$ essentially possess all these attributes, 
the {\it formal-concept property}
as the logical {\bf AND} of the members in $X^I$ 
faithfully characterises the 
FCL intent. 
Likewise, the {\it general} formal-concept property ({\it Gfcp})
is obtained via the
extension {from $F(G,M)$ to $F^\ast(G^\ast,M)$} as
%Similarly, after extending $M$ to $M^\ast$,
\begin{equation}
\eta(X)=\prod X^{I^\ast}=\prod \lbrace \mu \in M^\ast\mid\ \mu^R\supseteq X\rbrace=\prod_{X_0\supseteq X}\left(\prod [X_0]_F\right)\equiv \prod [X]_F.\label{eq:etaX}
\end{equation}
\item
Moreover, the conjugate relation $\eta(X^c)=\neg\rho(X)$ (Proposition~3.7~in~Ref.~\cite{LLJD12-1}) 
can have a natural
interpretation in terms of the conventional modal logics.
With $\rho(X)$, it is {\it not} possible that
any object $x$ in $X^c$ possess the property $\rho$ since 
otherwise $x\in X$. 
Hence, by \q{{\it not} possible $=$ {\it definitely} not}
any object $x\in X^c$ {\it definitely} has the property  
{\bf NOT} $\rho(X)$, also,  
any object in $X^c$ definitely possesses $\eta(X^c)$. 
Therefore, $\eta(X^c)=\neg \rho(X)$.
Note that in Eq.~(\ref{eq:rhoX}) and (\ref{eq:etaX})  
the operators 
{$I^\ast$ and $\Box^\ast$} are respectively obtained 
from {$I$ and $\Box$} by means of extending the attribute range 
to $M^\ast$, 
%{$\Diamond^\ast$}
and the same relationship also holds 
between $\Diamond^\ast$ and $\Diamond$ (Definition~3.3~in~Ref.~\cite{LLJD12-1}). 
\end{itemize}

It should however be clear that the emergence of {\bf GCL} 
need not be based on the RSL and FCL, although
the {\bf GCL} can be acquired as a common generalisation 
from RSL and FCL, see Lemma~2.8 and 3.1 of Ref.~\cite{LLJD12-1}.
Indeed, one can
use the content of {\bf GCL} 
to construct several extensions of the
RSL and FCL.
%which are subject to 
%$F^\ast(G,M^\ast)$.\\
%\begin{eqnarray}
%X\subseteq G &\mapsto& X^{I^\ast}=\left\lbrace \mu\in M^\ast \mid gR\mu,\ g \in X \right\rbrace\subseteq M^\ast,\nonumber\\
%Y\subseteq M^\ast &\mapsto& Y^{I^\ast}=\left\lbrace g \in G \mid gR\mu,\ \mu\in Y \right\rbrace\subseteq G,
%\nonumber\\ 
%X\subseteq G &\mapsto& X^{\Box^\ast}
%=\left\lbrace \mu\in M^\ast \mid \forall g\in G,\ gR\mu \Rightarrow g\in X \right\rbrace\subseteq M^\ast,\nonumber\\
%Y\subseteq M^\ast &\mapsto& Y^{\Box^\ast}
%=\left\lbrace g \in G \mid \forall \mu \in M^\ast,\ gR\mu \Rightarrow \mu\in Y \right\rbrace\subseteq G,\nonumber\\
%X\subseteq G &\mapsto& X^{\Diamond^\ast}=\lbrace \mu \in M^\ast \mid \exists g\in G, (gR\mu,\ g\in X) \rbrace\subseteq M^\ast, \nonumber\\
%Y\subseteq M^\ast &\mapsto& Y^{\Box^\ast}
%=\left\lbrace g \in G \mid \exists \mu \in M^\ast, (gR\mu,\ \mu\in Y) \right\rbrace\subseteq G.
%\label{eq:generalised_derivation}
%\end{eqnarray}
%{\bf Proposition~2.2}\\
\prp{
Subject to $F(G,M)$, 
a family of concept-lattice generalisations can be given  
%by means of the generalised operations 
%\q{$I$ to $I^\ast$, $\Box$ to $\Box^\ast$, $\Diamond$ to $\Diamond^\ast$}
%\q{$I^\ast$, $\Box^\ast$, $\Diamond^\ast$}
%(Definition~3.3  \cite{LLJD12-1}), which are thought of as the consequence 
%to employ $F^\ast(G,M^\ast)$, 
as follows.
\begin{itemize}  
\item
The generalised FCL ({gFCL}) can be accomplished by 
the {gFCL} concept
$(X,Y)_{\mbox{\tiny gFCL}}$ which satisfies $X^{I^\ast}=Y$ and $Y^{I^\ast}=X$, where 
$Y=\bigcup_{X_0\supseteq X}[X_0]_F\ \forall X\in E_F$. 
\item
The generalised RSL ({gRSL}) can be accomplished by the {gRSL} concept 
$(X,Y)_{\mbox{\tiny gRSL}}$
which satisfies $X^{\Box^\ast}=Y$ and $Y^{\Diamond^\ast}=X$, where 
$Y=\bigcup_{X_0\subseteq X}[X_0]_F\ \forall X\in E_F$. 
\item
The {\it complementary} generalised RSL ({cgRSL}) can be furnished by the {cgRSL} concept $(X,Y)_{\mbox{\tiny cgRSL}}$
which satisfies $X^{\Diamond^\ast}=Y$ and $Y^{\Box^\ast}=X$, 
where $Y=M^\ast\backslash \bigcup_{X_0\subseteq G\backslash X}[X_0]_F
\ \forall X\in E_F$.  
\end{itemize}
}
%{\bf Proof}:\\ 
\prf{
Note that the relations among 
the derivation operators 
$I,\Box$ and $\Diamond$ (Eq.~(4)-(7)~in~Ref.~\cite{LLJD12-1}) are preserved  
under the substitutions 
$\left\lbrace \begin{smallmatrix} 
I\rightarrow I^\ast\\
\Box\rightarrow\Box^\ast\\
\Diamond\rightarrow\Diamond^\ast
\end{smallmatrix}\right.$ since
$F^\ast(G,M^\ast)$ can be regarded as a new formal context.
\begin{itemize}
\item
%If $\nu\in \left\lbrace \mu\in M^\ast \mid gR\mu,\ g \in X_1 \right\rbrace$ for some 
%$X_1\supset X$, then $\nu\in \left\lbrace \mu\in M^\ast \mid gR\mu,\ g \in X \right\rbrace$. 
%Therefore, 
Upon employing Eq.~(9) of Ref.~\cite{LLJD12-1} with the 
generalisation {from $M$ to $M^\ast$}, 
\[
X^{I^\ast}=\left\lbrace \mu\in M^\ast \mid gR\mu,\ g \in X \right\rbrace
%=\left\lbrace \mu\in M^\ast \mid \mu^R=X_0, X_0\supseteq X \right\rbrace
=\bigcup_{X_0\supseteq X}\left\lbrace \mu\in M^\ast \mid \mu^R=X_0\right\rbrace=\bigcup_{X_0\supseteq X}[X_0]_F.
\]
Accordingly, consider $Y=X^{I^\ast}=\bigcup_{X_0\supseteq X}[X_0]_F$, where
\[
Y^{I^\ast}=(\bigcup_{X_0\supseteq X}[X_0]_F)^{I^\ast}=\bigcap_{X_0\supseteq X}([X_0]_F)^{I^\ast}=\bigcap_{X_0\supseteq X}\left\lbrace \mu\in M^\ast \mid \mu^R=X_0\right\rbrace^{I^\ast}=\bigcap_{X_0\supseteq X}X_0=X,
\]
namely, $Y^{I^\ast}=X$. %$\blacksquare$  
\item
Likewise, by Eq.~(9) of Ref.~\cite{LLJD12-1},
\[
X^{\Box^\ast}=\left\lbrace \mu\in M^\ast \mid \forall g\in X,\ gR\mu \implies g\in X \right\rbrace=\bigcup_{X_0\subseteq X}\left\lbrace \mu\in M^\ast \mid \mu^R=X_0\right\rbrace=\bigcup_{X_0\subseteq X} [X_0]_F=Y.
\]
Moreover, $Y^{\Diamond^\ast}=(\bigcup_{X_0\supseteq X}[X_0]_F)^{\Diamond^\ast}
=\bigcup_{X_0\subseteq X}([X_0]_F)^{\Diamond^\ast}=\bigcup_{X_0\subseteq X}X_0=X$. %$\blacksquare$
\item 
As the generalisation of result in Ref. \cite{YY04},
if $(X,Y)$ satisfies 
$\left\lbrace\begin{smallmatrix}
X^{\Box^\ast}=Y\\
Y^{\Diamond^\ast}=X
\end{smallmatrix}\right.$ 
then $\lsmatr{(X^c)^{\Diamond^\ast}=Y^c\\
(Y^c)^{\Box^\ast}=X^c}$, where
apparently $\lsmatr{X^c=G\backslash X\\
Y^c=M^\ast\backslash Y}$ since $X\subseteq G$ and $Y\subseteq M^\ast$.
%then $\left\lbrace\begin{smallmatrix}
%(G\backslash X)^{\Diamond^\ast}=M^\ast\backslash Y\\
%(M^\ast\backslash Y)^{\Box^\ast}=G\backslash X
%\end{smallmatrix}\right.$. 
%This is because $\left \lbrace\begin{smallmatrix}
%c\Box c=\Diamond\\
%c\Diamond c=\Box
%\end{smallmatrix}\right.$ 
%($\left\lbrace\begin{smallmatrix}
%S^c:=G\backslash S\ \mbox{for}\ S\subseteq G\\
%S^c:=M\backslash S\ \mbox{for}\ S\subseteq M
%\end{smallmatrix}\right.$)
%\cite{YY04}, hence $\left \lbrace\begin{smallmatrix}
%c\Box^\ast c=\Diamond^\ast\\
%c\Diamond^\ast c=\Box^\ast
%\end{smallmatrix}\right.$ where 
%$\left\lbrace\begin{smallmatrix}
%S^c:=G\backslash S\ \mbox{for}\ S\subseteq G\\
%S^c:=M^\ast\backslash S\ \mbox{for}\ S\subseteq M^\ast
%\end{smallmatrix}\right.$.
Then, the expression 
$\forall X\ (X^c,(\bigcup_{X_0\subseteq X}[X_0]_F)^c)_{\mbox{\tiny cgRSL}}$ 
can be used to construct a concept
lattice since expressions in this form are equipped with well-defined 
partial order.
Subsequently, because both $X$ and $X^c$ belong to $E_F$ and 
one prefers using the expression with respect to $X$, 
the cgRSL thus ranges over all the $X\in E_F$ with 
$(X,(\bigcup_{X_0\subseteq X^c}[X_0]_F)^c)_{\mbox{\tiny cgRSL}}\equiv 
(X,M^\ast\backslash (\bigcup_{X_0\subseteq G\backslash X}[X_0]_F))_{\mbox{\tiny cgRSL}}$. 
%$\blacksquare$
\end{itemize}
}
Certainly, on returning to the conventional scope based on 
$F(G,M)$,
the {gFCL} will be restricted to the FCL,
the {gRSL} will be restricted to the RSL, 
and the {cgRSL} will be restricted to 
the {\it property-oriented} RSL \cite{YY04}.
Nevertheless, while both the {gFCL} and {gRSL}
can be thought of as {\it object-oriented}, %in view of Ref. \cite{YY04}, 
the {cgRSL} here cannot be regarded as
a {\it property-oriented} lattice 
that generalises the {\it property-oriented} RSL.
%This is because the {cgRSL}
%is not employing $G$ as the formal attributes
%for categorising $M^\ast$ (cf. Table \ref{table:formal_scheme}). 

Notably,
apart from %all the concept lattices in 
{\bf Proposition~2.2},
still more concept lattices could have been generated from the {\bf GCL} by means of the general intents.
For instance, one could also consider
$(X,Y)_{\mbox{new}}=(X,M^\ast\backslash (\bigcup_{X_0\supseteq X}[X_0]_F))\ \forall X\in E_F$, where $Y$ is the complementary set of the {gFCL} intent.
However, for all those constructions it remains
fundamental to have the intelligibility 
about why to relate the object class and its associating property
in particular ways.
%than to achieve the Galois connection required by 
%constructing a lattice. 
It is based on such intelligibility
that the analytics 
accompanied with lattice can be performed.
In this regard, the {gFCL} deals with
the {\it necessary} feature
an object in any given class should exhibit
and the {gRSL} the {\it sufficient} feature 
by which an object can be categorised into 
a definite class, while the artificial 
constructions like {cgRSL} or $(X,Y)_{\mbox{new}}$ could become less significant. 
%since their intelligibility  
%is as yet unobvious.
Another point is that
the lattices introduced in {\bf Proposition 2.2}
are {\it nested} in the sense 
that the attributes in $M^\ast$ are used as intents in a repeated manner.
Clearly, the fact that the intents overlap 
can render the analysis difficult.
Nevertheless, both the {gFCL} and the {gRSL} can be regarded as 
a {\it half} of the {\bf GCL}, whose intents are then disjoint ({\bf Proposition~2.1}). 
For concreteness, if $(X, Y_1)$ is a {gFCL} concept
and $(X, Y_2)$ is a {gRSL} concept then $(X,Y_1\cap Y_2)$
is a {\it general concept}.
%Of particular significance is also that 
%the {\bf GCL} 
%categorises 
%whatever object classes one can differentiate
%exhausts  
%whatever attributes one can construct out of $M$ for the general intents
%in the manner that 
%every attribute entity in $M^\ast$ is 
%only employed {\it once}.
The tractability problem for the 
{\bf GCL} content will then be resolved
as follows. 
\begin{itemize}
\item For the lattice construction:\\
%For the construct
The general extents 
are known in advance subject to $F(G,M)$, which are the 
$2^{n_F}$ members in $E_F=\lbrace\mu^R\mid \mu\in M^\ast\rbrace=\sigma (\lbrace m^R\mid m\in M\rbrace)$. 
The generalised attribute-set $M^\ast$ 
is distributed to the $2^{n_F}$ nodes as general intents, each of which
is expressible in terms of a closed interval.
Thus, one only needs the $2\times 2^{n_F}$ bounds 
({\bf Proposition~2.1}) for fixing down the general intents,
which further reduces to $2^{n_F}$ attributes since 
$\forall X\in E_F\ \eta(X^c)=\neg\rho(X)$ 
and $X \in E_F\iff X^c \in E_F$ (Proposition~3.7~in~Ref.~\cite{LLJD12-1}).
Basically, these $2^{n_F}$ attributes can be determined 
using a fundamental $n_F$ attribute construction (Proposition~3.9, 3.11 and 3.12; Corollary~3.13~in~Ref.~\cite{LLJD12-1}).
However, it will be further shown in the coming section
that they can be directly read out from the formal context;
the practical construction for {\bf GCL} 
is as tractable as {\it listing out the formal context}. 
\item For the implication relations:\\
%For the logical deduction developed from {\bf GCL}:\\
The {\bf GCL} supports
the logic deduction 
by characterising any object class 
of interest in terms of 
{\it non-overlapping} general intent, 
whose upper and lower bounds corresponds to
the {\it sufficient} and {\it necessary} properties of 
the given object class, respectively. 
To be concrete, 
attributes grouped into the same general intent 
are {\it logically equivalent}
since they correspond to the property of the same object class.
One will show in Sec. \ref{four} that
a unique formula based on the contextual truth $1_\eta$ or 
falsity $0_\rho$
(Proposition~3.12~in Ref.~\cite{LLJD12-1}) 
suffices to generate all 
such logic implications established between 
any attribute pairs in $M^\ast$.
%the set-inclusion relation between two general extents
%induces the {\it logic implications} between members of 
%the two corresponding intents and implies 
%that the property of an object set
%is essentially a part of the feature of its super set.
%In particular,
%As for the reason 
%why such extreme simplicity may occur, it can be understood 
%that the general intents are disjoint in the {\bf GCL} construction, all 
%the attributes will not be employed twice, therefore, the complexity is reduced.
\end{itemize}

\begin{table}
\begin{tabular}{|p{0.6cm}|c|c|c|c|c|} 
\multicolumn{1}{r}{$F(G,M)$}&\multicolumn{5}{r}{}\\
\hline
 \multicolumn{1}{|r|}{\footnotesize{$M$}}  &\multirow{2}{*}{$a$}&\multirow{2}{*}{$b$}&\multirow{2}{*}{$c$}&\multirow{2}{*}{$d$}&\multirow{2}{*}{$e$}\\ 
\multicolumn{1}{|l|}{\footnotesize{$G$}}   &{}                  &{}                  &{}                  &{}                  &{}\\ \hline
   {$\ \ 1$}     &$\times$&{}       &$\times$&$\times$&$\times$ \\ \hline
   {$\ \ 2$}     &$\times$&{}       &$\times$&{}      &{}       \\ \hline
   {$\ \ 3$}     &{}      &$\times$ &{}      &{}      &$\times$ \\ \hline
   {$\ \ 4$}     &{}      &$\times$ &{}      &{}      &$\times$  \\ \hline
   {$\ \ 5$}     &$\times$&{}       &{}      &{}      &{}       \\ \hline
   {$\ \ 6$}     &$\times$&$\times$ &{}      &{}      &$\times$ \\ \hline
\end{tabular}\quad\quad
\begin{tabular}{|p{0.6cm}|c|c|c|c|c|c|} 
\multicolumn{1}{r}{$F^\prime(G,M)$}&\multicolumn{6}{r}{}\\
\hline
 \multicolumn{1}{|r|}{\footnotesize{$M$}}  &\multirow{2}{*}{$1$}&\multirow{2}{*}{$2$}&\multirow{2}{*}{$3$}
&\multirow{2}{*}{$4$}&\multirow{2}{*}{$5$}&\multirow{2}{*}{$6$}\\ 
\multicolumn{1}{|l|}{\footnotesize{$G$}}   &{}                  &{}                  &{}                  &{}         &{}         &{}\\ \hline
   {$\ \ a$}     &$\times$&$\times$       &{}&{}&$\times$ &$\times$\\ \hline
   {$\ \ b$}     &{} &{}       &$\times$&$\times$      &{}      &$\times$ \\ \hline
   {$\ \ c$}     &$\times$      &$\times$ &{}      &{}      &{} &{}\\ \hline
   {$\ \ d$}     &$\times$       &{} &{}      &{}      &{}&{}  \\ \hline
   {$\ \ e$}     &$\times$&{}       &$\times$&$\times$      &{}      &$\times$ \\ \hline
\end{tabular}
\caption{The formal context
$F(G, M)$ with 
$(G,M)=(\lbrace 1,2,3,4,5,6\rbrace,\lbrace a,b,c,d,e\rbrace)$ and
the formal context $F^\prime(G, M)$ with $(G,M)=(\lbrace a,b,c,d,e\rbrace,\lbrace 1,2,3,4,5,6\rbrace)$\\ 
Although transposing one of the contexts gives rise to the other one, 
there are two mathematically distinct concept lattices to be 
constructed.
 %can be explained as follows. 
One may assume $U=\lbrace 1,2,3,4,5,6\rbrace$ 
and $V=\lbrace a,b,c,d,e\rbrace$ 
to be what are conventionally 
recognised as objects and attributes, respectively.
According to the setting of Ref. \cite{YY04}, 
the choice $F(G, M)$ in which $(U,V)\equiv (G,M)$ 
is referred to as an object-oriented scheme
and $F^\prime(G, M)$ ($(V,U)\equiv (G,M)$) a property-oriented scheme.}
\label{table:formal_scheme}
\end{table}
\begin{table}   
\begin{tabular}{|p{1cm}|c|c |c|c |c|c |c|c |c|c |c|c |c|c |c|c |ccc|} \cline{2-20}
\multicolumn{1}{c}{}&\multicolumn{19}{|c|}{$\longleftarrow\quad\quad 4294967296$ columns grouped into $32$ equivalent classes $\quad\quad \longrightarrow$}\\ 
\hline
\multicolumn{1}{|r|}{\multirow{2}{*}{$\left[X\right]_F$}}
&\multicolumn{2}{|c|}{\multirow{2}{*}{$\left[\left\lbrace 1,2,5,6\right\rbrace\right]_F$}}
&\multicolumn{2}{|c|}{\multirow{2}{*}{$\left[\left\lbrace 3,4,6\right\rbrace\right]_F$}}
&\multicolumn{2}{|c|}{\multirow{2}{*}{$\left[\left\lbrace 1,2\right\rbrace\right]_F$}}
&\multicolumn{2}{|c|}{\multirow{2}{*}{$\left[\left\lbrace 1\right\rbrace\right]_F$}}
&\multicolumn{2}{|r|}{\multirow{2}{*}{$\left[\left\lbrace 1,3,4,6\right\rbrace\right]_F$}}
&\multicolumn{2}{|c|}{\multirow{2}{*}{$\left[\emptyset\right]_F$}}
&\multicolumn{2}{|c|}{\multirow{2}{*}{$\left[G\right]_F$}}
&{\multirow{2}{*}{$\left[\left\lbrace 2\right\rbrace\right]_F$}}
&{\multirow{2}{*}{$\left[\left\lbrace 3,4\right\rbrace\right]_F$}}
&\multicolumn{3}{|c|}{\multirow{2}{*}{etc.}}\\ 
\multicolumn{1}{|r|}{}
&\multicolumn{2}{|c|}{}
&\multicolumn{2}{|c|}{}
&\multicolumn{2}{|c|}{}
&\multicolumn{2}{|c|}{}
&\multicolumn{2}{|r|}{}
&\multicolumn{2}{|c|}{}
&\multicolumn{2}{|c|}{}
&{}
&{}
&\multicolumn{3}{|c|}{}\\ 
\hline  
 \multicolumn{1}{|r|}{{$\ M^\ast$}}  
&\multirow{2}{*}{$a$}&\multirow{2}{*}{$\ldots$}
&\multirow{2}{*}{$b$}&\multirow{2}{*}{$\ldots$}
&\multirow{2}{*}{$c$}&\multirow{2}{*}{$\ldots$}
&\multirow{2}{*}{$d$}&\multirow{2}{*}{$\ldots$}
&\multirow{2}{*}{$e$}&\multirow{2}{*}{$\ldots$}
&\multirow{2}{*}{${\bf 0}$}&\multirow{2}{*}{$\ldots$}
&\multirow{2}{*}{${\bf 1}$}&\multirow{2}{*}{$\ldots$}
&\multirow{2}{*}{$\ldots$}&\multirow{2}{*}{$\ldots$}
&\multirow{2}{*}{$\ldots$}&\multirow{2}{*}{$\ldots$}
&\multirow{2}{*}{$\ldots$}
\\ 
\multicolumn{1}{|l|}{\footnotesize{$G_{/R}$}}&{}&{}  &{}      &{}             &{}      &{}             &{}      &{}             &{}&{}
&{}&{}&{}&{}&{}&{}&{}&{}&{}\\ 
\hline
   {$\ \ \left\lbrace1\right\rbrace$}     &$\times$&$\ldots\times$            &{}      &{}             &$\times$&$\ldots\times$ &$\times$&$\ldots\times$ &$\times$&$\ldots\times$ 
&{}&{}&$\times$&$\ldots\times$&{}&{}&$\ldots$&$\ldots$&$\ldots$ 
\\ \hline
   {$\ \ \left\lbrace2\right\rbrace$}     &$\times$&$\ldots\times$            &{}      &{}             &$\times$&$\ldots\times$ &{}      &{}             &{}&{}       
&{}&{}&$\times$&$\ldots\times$&$\times\ldots\times$&{}&$\ldots$&$\ldots$&$\ldots$ 
\\ \hline
   {$\  \left\lbrace3,4\right\rbrace$}    &{}      &{}                        &$\times$&$\ldots\times$ &{}      &{}             &{}      &{}            &$\times$&$\ldots\times$ 
&{}&{}&$\times$&$\ldots\times$&{}&$\times\ldots\times$&$\ldots$&$\ldots$&$\ldots$ 
\\ \hline
   {$\ \ \left\lbrace5\right\rbrace$}     &$\times$&$\ldots\times$            &{}      &{}             &{}      &{}             &{}      &{}             &{}&{}       
&{}&{}&$\times$&$\ldots\times$&{}&{}&$\ldots$&$\ldots$&$\ldots$ 
\\ \hline
   {$\ \ \left\lbrace6\right\rbrace$}     &$\times$&$\ldots\times$            &$\times$&$\ldots\times$ &{}      &{}             &{}      &{}             &$\times$&$\ldots\times$ 
&{}&{}&$\times$&$\ldots\times$&{}&{}&$\ldots$&$\ldots$&$\ldots$ 
\\ \hline
\end{tabular}
\caption{The factorisation of $|M^\ast|$ attributes into $2^{n_F}$ equivalent classes, exemplified with $F(G, M)$ in
Table \ref{table:formal_scheme}\\
For convenience, it is assumed that all the members in $M$ are independent. 
Based on a given formal context 
it is possible, albeit tedious, to determine
the object correspondence to
all the attributes 
that can be constructed out of $M$, i.e. $M^\ast $
because this is nothing but the completion of a truth-value table.
Accordingly, any $\mu$ in $M^\ast$ must end up with $\mu^R\in E_F$ 
($\mu^R\subseteq G$). 
It turns out that $M^\ast $ can be further categorised:
since $n_F=5$ ($|E_F|=2^{|n_F|}$), the $|M^\ast|= 2^{2^5}=4294967296$ generalised 
attributes can be grouped into $2^5$ equivalent classes ($[X]_F$s). 
Indeed, $|[X]_F|$ is even a constant over the {\bf GCL} ({\bf Corollary~4.4}).
%It should be still reminded "that $n_F$ happens to be $|M|$ is an accident". In general, $n_F$
%ranges over the closed interval $[1, 2^{|M|}]$. 
}
\label{table:equivalent_class} 
\end{table}

\section{lattice construction per read out}\label{three}
\begin{figure}   
\includegraphics[scale=0.3,angle=0]{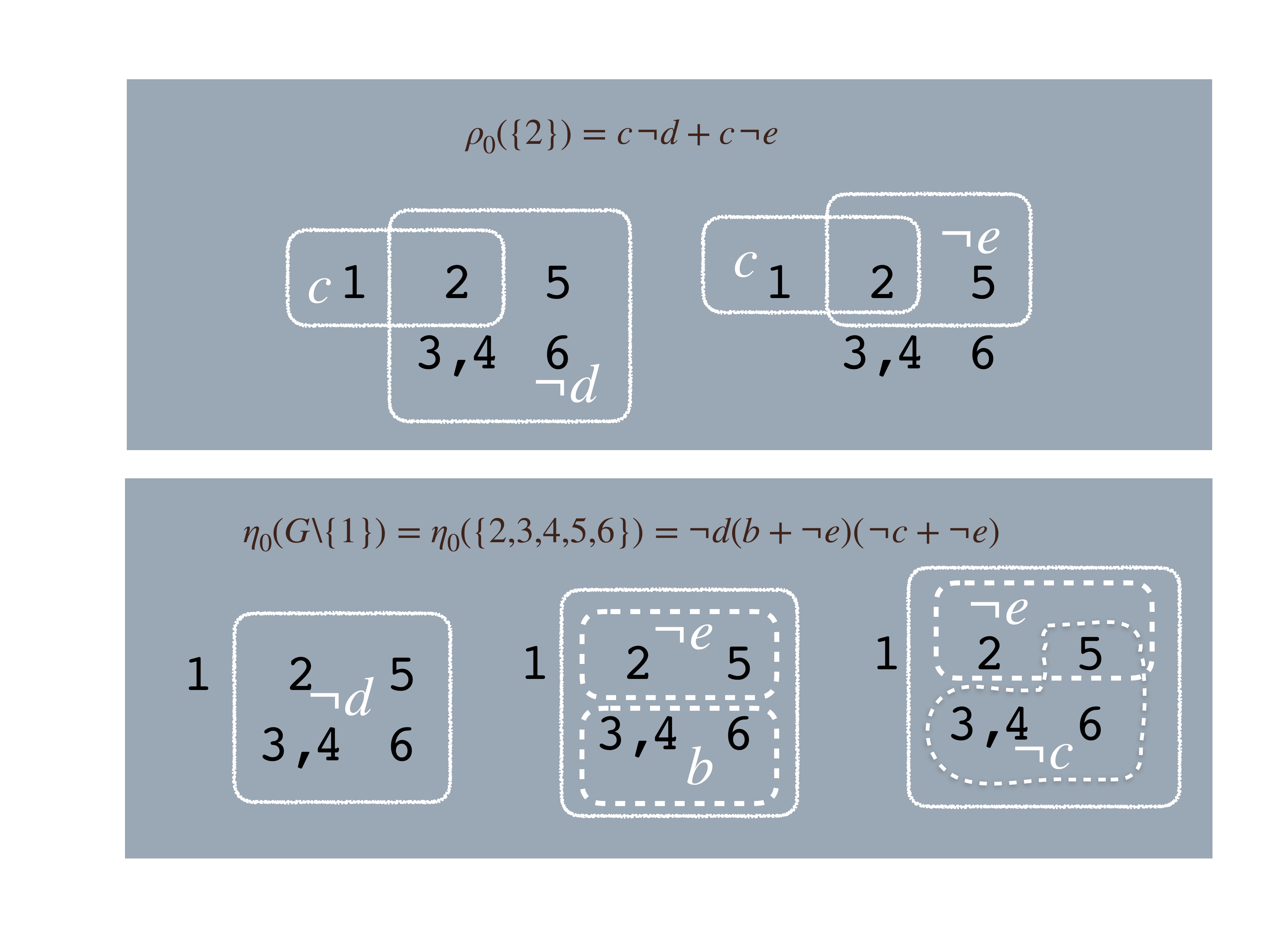}
\caption{
Obtaining $\rho_0(D_k)$ and $\eta_0(G\backslash D_k)$, via the {\it irreducibility} requirements from Eq.~(\ref{eq:irreducible_det}), for the case of Fig.~\ref{figure:context_venn}\\
Resolving in this manner needs not be the most practical way to construct {\bf GCL}, but it provides an intuitive understanding for the {\it irreducibility}. 
$\rho_0(\lbrace 2\rbrace)=\sum [\lbrace 2\rbrace^\times]_F=c\neg d+c\neg e$
in the sense that $c\neg d$ and $c\neg e$ are {\it irreducible} conjunctions
which render $c^R\cap (\neg d)^R=c^R\cap (\neg e)^R=\lbrace 2 \rbrace$.
Similarly, $\eta_0(G\backslash\lbrace 1\rbrace)=
\prod [\lbrace 2,3,4,5,6\rbrace^+]_F
=\neg d (b+\neg e)(\neg c+\neg e)$
since $\neg d, (b+\neg e)$ and $(\neg c+\neg e)$ are {\it irreducible} disjunctions for which $\neg d^R=(b+\neg e)^R=(\neg c+\neg e)^R=\lbrace 2,3,4,5,6\rbrace$.} 
\label{fig:construct_irreduicble}
\end{figure}

%The basic constituents for the 
%{\bf GCL} are the smallest object-sets discernible by the formal context, say $G_{/R}:=\lbrace D_k\mid 1\leq k\leq n_F\rbrace$.
% \footnote{$D_k$ has been named an $F$-disjoint subset after \cite{LLJD12-1} as it behaves like a disjoint set subject to the formal context $F$.}.
To proceed with the construction 
of the general concept ({\bf Proposition~2.1})
\begin{equation}
(X,\rho(X),\eta(X))\equiv(X,[X]_F)=(X,[\eta(X),\rho(X)]),
\label{eq:g_concept_23}
\end{equation}
%it is instructive 
%to re-examine the method proposed in the first paper of this series, the main purpose of which 
%is to make the {\bf GCL} comparable with the RSL and FCL.
%For convenience, 
let
$X=\bigcup_{D_k\subseteq X} D_k$ and $X=\bigcap_{G\backslash D_k\supseteq X}G\backslash D_k$ for any general extent, where $D_k$'s %belong to $G_{/R}$, which 
are 
the smallest object sets discernible by the formal context 
(Definition~2.4, Proposition~3.5~in Ref.~\cite{LLJD12-1}).
%With such an expression
%According to
After Proposition~3.7~and~3.9 of Ref.~\cite{LLJD12-1},
%$\forall X\in E_F$ 
%$\rho(X)=\sum \lbrace \rho(D_k)\mid D_k\subseteq X\rbrace$ 
%and $\eta(X)=\prod \lbrace \eta(G\backslash D_k)\mid G\backslash D_k\supseteq X\rbrace$.
one may end up with the relations  
\begin{eqnarray}
\eta(X)=\prod \lbrace \eta(G\backslash D_k)\mid G\backslash D_k\supseteq X\rbrace, & & 
\rho(X)=\sum \lbrace \rho(D_k)\mid D_k\subseteq X\rbrace,
\nonumber \\
\eta(X)=\eta_0(X)\cdot 1_\eta,& & \rho(X)=\rho_0(X)+0_\rho, \nonumber \\
\eta(G)\equiv 1_\eta=\prod [G^{+}]_F,& & \rho(\emptyset)\equiv 0_\rho=\sum [\emptyset^{\times}]_F,\nonumber \\
\eta_0(G\backslash D_k)=\prod [(G\backslash D_k)^{+}]_F,& & \rho_0(D_k)=\sum [D_k^{\times}]_F,\label{eq:irreducible_det} 
\end{eqnarray}
where $X^+$ ($X^\times$) is referred to as an $X$-irreducible disjunction (conjunction) (Definition~2.14, Corollary~3.13~in Ref.~\cite{LLJD12-1}).
In addition, $\rho(\emptyset)=0_\rho$ and $\eta(G)=1_\eta$
since $\rho(\emptyset)$ plays 
the role of falsity for all the {\it upper} bounds
and $\eta(G)$ plays the role of truth for all the {\it lower} bounds 
(Proposition~3.12~in Ref.~\cite{LLJD12-1}). 
%This is in effect the determination starting with 
%the fundamental attribute-collections 
%${\cal P}_\eta=
%\lbrace \eta(G), \eta(G\backslash D_1),\ldots \eta(G\backslash D_{n_F})\rbrace
%\equiv \lbrace 1_\eta, \eta_0(G\backslash D_1)\cdot 1_\eta,\ldots \eta_0(G\backslash D_{n_F})\cdot 1_\eta\rbrace$
%and
%${\cal P}_\rho
%=\lbrace \rho(\emptyset), \rho(D_1),\ldots \rho(D_{n_F})\rbrace
%\equiv
%\lbrace 0_\rho, \rho_0(D_1)+0_\rho,\ldots \rho_0(D_{n_F})+0_\rho\rbrace$.
%Accordingly, one may \footnote{To implement Eq. \ref{eq:irreducible_det} as computation algorithms is in fact straightforward.
%However, doing this right here could digress the approach at hand.} 
%explicitly determine all such fundamental attributes 
%for {\it Grsp} and for {\it Gfcp}. 
%\nonumber
As an example, Fig. \ref{fig:construct_irreduicble} illustrates how  ${\rho_0(D_k)}$ and 
${\eta_0(G\backslash D_k)}$ 
are determined from the irreducible expressions 
given in Eq.~(\ref{eq:irreducible_det}).
Specifically, in the same manner,
\begin{eqnarray}
\eta (G)\equiv 1_\eta=\prod [G^{+}]_F&=&
(a+b)(a+e)(a+\neg c)(a+\neg d)(b+c+\neg d)(b+c+\neg e)(b+d+\neg e)\nonumber\\
&& (b+\neg d+e)(\neg b+\neg c)(\neg b+\neg d)(\neg b+e)(c+\neg d) (\neg c+d+\neg e)(\neg d+e),\label{eq5}
\end{eqnarray}
which conjuncts all the {\it irreducible disjunctions} of the attributes in 
$\breve M=M\cup\lbrace \neg m\mid m\in M\rbrace$. 
Notably, $$(a+b)^R=\ldots=(b+d+\neg e)^R
=(b+\neg d+e)^R=\ldots=(\neg d+e)^R=G,$$
where, e.g., $(b+c+\neg d)\in [G^{+}]_F$ is irreducible
because $(b+c+\neg d)^R=G$ but 
none of $\lrbr{(b+c)^R,(c+\neg d)^R,(b+\neg d)^R}$ can be identified with $G$. 
%$\lsmatr{(b+c)^R\neq G\\
%(c+\neg d)^R\neq G\\
%(b+\neg d)^R\neq G
%}$.
%Likewise, 
%\begin{equation}
%\rho(\lbrace 5\rbrace)=\sum [\lbrace 5\rbrace^\times]_F=\neg b\neg c+ \neg c\neg e \label{eq7}
%\end{equation} 
%subject to the "irreducibility" that $(\neg b\neg c)^R=(\neg c\neg e)^R=\lbrace 5\rbrace$ 
%but $\neg b^R\neq \lbrace 5\rbrace, \neg c^R\neq \lbrace 5\rbrace$ and $\neg e^R\neq \lbrace 5\rbrace$.

Note that the approach based on Eq. (\ref{eq:irreducible_det}) 
in general renders $\rho(X)$ in {\bf DNF} and $\eta(X)$
in {\bf CNF}, which is instructive for the generality of {\bf GCL}
in relation to the other lattices
since one has in effect obtained $\rho(X)$ in the style of RSL
and $\eta(X)$ in the style of FCL.
However, such an approach is rather tedious
thus should not be recommeded in actual practice.   
%one reverts such convention and shows that it 
%potentially leads to the binary-mask representation of {\bf GCL}.
%In fact, based on the self-duality (Lemma 3.7 \cite{LLJD12-1})  
%either of ${\cal P}_\eta$ and ${\cal P}_\rho$ will suffice the {\bf GCL}-construction; let one start with
%${\cal P}_\eta$ and consider $\lbrace 1_\eta, \eta_0(G\backslash D_1),\ldots \eta_0(G\backslash D_{n_F})\rbrace$
%.\\  
Now one proceeds to show 
that adopting $\eta(X)$ in {\bf DNF} and $\rho(X)$ in {\bf CNF}
will provide a simpler construction
which potentially leads to the full determination of general concepts per read out.
%{\bf Proposition~3.1}\\*
\prp{
Subject to a formal context $F(G,M)$,
$\eta (G)\equiv 1_\eta=
\sum_{k=1}^{n_F}\eta(D_k)$.
In effect, $1_\eta$ is obtained by summing up all the 
lower bounds of intents corresponding to disjoint regions on 
the contextual Venn diagram, cf. Fig.~\ref{figure:context_venn}. 
}  
%$\rho (\emptyset)\equiv 0_\rho$  in {\bf CNF} is $\prod_{g\in G}\prod g^{\breve R}\equiv \sum_{k=1}^{n_F}\prod D_k^{\breve I}=
%\sum_{k=1}^{n_F}\eta(D_k)$  
%{\bf Proof}:\\
%Apparently, $\eta(D_k)=\prod D_k^{\breve I}$ as it coincides with Lemma 3.10 of \cite{LLJD12-1}.
%Hence, it remains to show $\eta (G)=\sum_{k=1}^{n_F}\eta(D_k)$.
\prf{
$\forall k\in [1, n_F]\ \eta(G) > \eta(D_k)$ by Proposition~3.10,~3.14 in Ref.~\cite{LLJD12-1}
$\therefore \eta (G)\geq \sum_{k=1}^{n_F}\eta(D_k)$.
%$\eta(G) =\prod [G]_F=\prod \lrbr{\mu\in M^\ast\mid\mu^R=G}$\\
%$\eta(D_k) =\prod [D_k]_F=\prod \lrbr{\mu\in M^\ast\mid\mu^R=D_k}$
On the other hand, $(\sum_{k=1}^{n_F}\eta(D_k))^R=\bigcup_{k=1}^{n_F}\eta(D_k)^R=G\ \therefore\sum_{k=1}^{n_F}\eta(D_k)\in [G]_F$.
Hence, $\eta (G)\leq \sum_{k=1}^{n_F}\eta(D_k)$ since $\eta(G)=\prod[G]_F$ is the lower bound for $[G]_F$ ({\bf Proposition~2.1}).} 
%%$\blacksquare$
%For convenience, extend the map $I$ to $\breve I$ 
%such that it can be applied to 
%$\breve M=M\cup\lrbr{\neg m\mid m\in M}$ as
%\begin{equation}
%X^{\breve I}
%=\lrbr{m\mid m\in g^R,\ g \in X }\cup \lrbr{\neg m\mid m\not \in g^R,\ g \in X }
%=\left\lbrace {\breve m}\in {\breve M} \mid gR{\breve m},\ g \in X \right\rbrace\subseteq {\breve M}.\label{Eq:breveI}
%\end{equation}
%\[
%{\breve R}: g\rightarrow {\breve Y}=g^{\breve R},\ g \in G,\ {\breve Y}\subset {\breve M}. 
%(R: {\breve m}\rightarrow X= {\breve m}^R,\ {\breve m} \in {\breve M}, X\subset G).
%\]
%Here, $D_k^{\breve I}=\lbrace m\in M \mid m\in g^R,\ g\in D_k\rbrace\cup \lbrace \neg m \mid m\not \in g^R,\ m\in M,\ g\in D_k\rbrace \in {\bf \Psi}_M$ (Definition 2.6 \cite{LLJD12-1}) is not ambiguous since all $g^R$s are equivalent within $D_k$.
Since $\eta(D_k)=\prod \Psi^k$, 
where $\Psi^k=\lbrace m\in M \mid m\in D_k^I\rbrace\cup 
\lbrace \neg m \mid m\not\in D_k^I, m\in M\rbrace$ by Proposition~3.10 in Ref.~\cite{LLJD12-1},
Eq. (\ref{eq5}) can also be computed as 
\begin{equation}
1_\eta =
\sum_{k=1}^{n_F} \Psi^k=
a\neg bcde+a\neg bc\neg d\neg e+\neg ab\neg c\neg de+a\neg b\neg c\neg d\neg e+ab\neg c\neg de.\label{eq:etadnf}
\end{equation}
%In addition, for all $g \in D_k$, one defines $D_k^{\breve I}=g^{\breve R}$ since all of objects in $D_k$ map onto 
%the same attribute-subset of ${\breve M}$.\\
%The promised simplification for constructing {\bf GCL} is based on
%$\eta(D_{k})\cdot\eta(D_{k^\prime})={\bf 0}$ iff $k\neq k^\prime$.
%Hence, $\mu\cdot \eta(D_{k})\in\lbrace \eta(D_{k}),{\bf 0}\rbrace$.
%This is because $\eta (D_k)$ is a non-trivial infimum for $M^\ast$ 
%(Lemma 2.12 \cite{LLJD12-1}), 
%which could not be modified by any further conjunction unless the result becomes trivial. 
%In other words, $\forall k,\ \eta (D_k)=\prod \Psi_M^k$ is an $M^\ast$- {\it atom} meaning that nothing lies between it and ${\bf 0}$
%(Lemma 3.10 \cite{LLJD12-1}).\\ 
%{\bf Lemma~3.2}\\
\lem{
Given a formal context $F(G,M)$, it can be shown that
\begin{itemize}
\item
$\eta(G\backslash D_k)\equiv \eta_0(G\backslash D_k)\cdot 1_\eta=\sum_{j\neq k}\eta(D_j)$,
\item
$\rho(D_k)\equiv \rho_0(D_k)+0_\rho=\prod_{j\neq k}\rho(G\backslash D_j)$.
\end{itemize}}
\prf{
%{\bf Proof}:\\
The fact that 
$\eta(G\backslash D_k)= \eta_0(G\backslash D_k)\cdot 1_\eta$
and $\rho(D_k)= \rho_0(D_k)+0_\rho$ is the direct consequence of 
Eq.~(\ref{eq:irreducible_det}).
\begin{itemize}
\item
%Alternatively, \q{$\eta(G\backslash D_k)=\sum_{j\neq k}\eta(D_j)$} can be achieved as follows.
$\eta (G\backslash D_k)\geq \sum_{j\neq k}\eta(D_j)$ since $\eta (G\backslash D_k)\geq \eta(D_j)$ by $G\backslash D_k\supseteq \eta(D_j)$ $\forall j\neq k$ due to Proposition~3.14~in Ref.~\cite{LLJD12-1}.
On the other hand, $\eta (G\backslash D_k)\leq \mu\ \forall\mu \in [G\backslash D_k]_F$, implying that 
$\eta (G\backslash D_k)\leq \sum_{j\neq k}\eta(D_j)$ by 
$(\sum_{j\neq k}\eta(D_j))^R=G\backslash D_k$. 
Therefore, $\eta (G\backslash D_k)=\sum_{j\neq k}\eta(D_j)$.
%By {\bf Lemma 3.2}, 
%$\eta_0(G\backslash D_k)\cdot \eta(D_k)={\bf 0}$. 
%Moreover, $\prod (D_j^{\breve I}\backslash D_k^{\breve I})=\prod \lbrace \breve m\in {\breve M} \mid {\breve m}\not\in D_k^{\breve I},\ {\breve m}\in D_j^{\breve I} \rbrace\geq \prod D_j^{\breve I}=\eta(D_j)\ \forall j\neq k$. 
%Therefore, $\eta_0(G\backslash D_k)\cdot 1_\eta=\sum_{j\neq k}\eta(D_j)$
%since $1_\eta=\sum_{k=1}^{n_F}\eta(D_k)$ ({\bf Propositions 3.1}). 
%$\blacksquare$ 
\item 
By the conjugateness of Proposition~3.7~in Ref.~\cite{LLJD12-1}, $\rho(D_k)=\neg \eta(G\backslash D_k)=\neg \sum_{j\neq k}\eta(D_j)$.
It follows then $\neg \sum_{j\neq k}\eta(D_j)=\prod_{j\neq k}\neg\eta(D_j)=\prod_{j\neq k}\rho(G\backslash D_j)$. 
%$\blacksquare$ 
\end{itemize}
}
%To reconfirm the above result, 
Let $D_1=\lbrace 1\rbrace$, $D_2=\lbrace 2\rbrace$, $D_3=\lbrace 3,4\rbrace$, 
$D_4=\lbrace 5\rbrace$ and $D_5=\lbrace 6\rbrace$.
In Fig.~\ref{fig:construct_irreduicble},
it can be checked that
\begin{eqnarray}
\eta_0(G\backslash D_1)\cdot 1_\eta&=&\neg d (b+\neg e)(\neg c+\neg e)\cdot 
1_\eta\nonumber\\
&=&a\neg bc\neg d\neg e+\neg ab\neg c\neg de+a\neg b\neg c\neg d\neg e+ab\neg c\neg de=\eta(G\backslash D_1).\nonumber
\end{eqnarray}
%Meanwhile, an equivalent expression is $X=\bigcap_{k\in K_0\backslash K} G\backslash D_k$. 
%Note these are precursors of binary representations for $\lbrace \mu^R\mid \mu\in M^\ast\rbrace$.  
%$X$ can be represented by a binary string 
%while $K\ (K_0\backslash K)$ is employed to note the positions at which "1" ("0") occur.
%All the {\it GI}s for the {\bf GCL} can be determined via their upper and lower bounds in the following manner.\\ 
%{\bf Proposition~3.3}\\
\prp{
Given a formal context $F(G,M)$, 
one may express the general extent in terms of 
$X=\bigcup_{k\in K} D_k\equiv \bigcap_{k\in K_0\backslash K} G\backslash D_k$ for some $K\subseteq K_0$, where 
the index set $K_0:=\lbrace 1,\cdots,\ n_F\rbrace$.
The upper and lower bounds for the corresponding general intents are as follows.
\begin{itemize}
\item
$\eta(X)=\sum_{k\in K}\eta(D_k)$ and $\rho(X)
=\prod_{k\in K_0\backslash K}\rho(D_k^c)$, which is
in contrast to Proposition~3.9 of Ref.~\cite{LLJD12-1}
that can be restated as
$\rho(X)=\sum_{k\in K}\rho(D_k)$ and $\eta(X)
=\prod_{k\in K_0\backslash K}\eta(D_k^c)$.
\item 
$\left\lbrace\begin{array}{ccc}
\eta(\bigcup_i X_i)=\sum_{i}\eta(X_i),&&\rho(\bigcap_j X_j)=\prod_{j}\rho(X_j)\\
\rho(\bigcup_i X_i)=\sum_{i}\rho(X_i),&&\eta(\bigcap_j X_j)=\prod_{j}\eta(X_j)
\end{array}\right.$
%$\eta(\bigcup_i X_i)=\sum_{i}\eta(X_i)$ and $\rho(\bigcap_j X_j)=\prod_{j}\rho(X_j)$, 
%$\eta(\bigcup_i X_i)=\sum_{i}\eta(X_i)$ and $\rho(\bigcap_j X_j)=\prod_{j}\rho(X_j)$,
in which $X_i,X_j\in E_F$.
\end{itemize}
} 
%{\bf Proof}:
\prf{
\begin{itemize}
\item
For $X\in E_F$,
$X=\bigcup_{D_k\subseteq X} D_k=\bigcap_{D_k^c\supseteq X}G\backslash D_k$, 
as is given in Proposition~3.5~of Ref.~\cite{LLJD12-1}, can be rewritten as 
$X=\bigcup_{k\in K} D_k=\bigcap_{k\in K_0\backslash K} G\backslash D_k$ for some $K\subseteq K_0$, where $G=\bigcup_{k\in K_0} D_k$. Hence,
%For $X=G$, it is obvious that
%$\eta (G)=\sum_{k=1}^{n_F}\eta(D_k)\equiv\sum_{k\in K_0}\eta(D_k)$ 
%({\bf Proposition 3.1}).
%In general, 
%$\prod_{k\in K_0\backslash K}\eta( G\backslash D_k)$ (Lemma~3.9~in Ref.~\cite{LLJD12-1}).
\[
\eta(X)=\prod_{k\in K_0\backslash K}\eta(D_k^c)
=\prod_{k\in K_0\backslash K}\sum_{k^\prime\neq k}
\eta(D_{k^\prime})
=\sum_{{k^\prime}\in \bigcap_{k\in K}{K_0\backslash \lbrace k\rbrace}}\eta(D_{k^\prime})
=\sum_{k\in K}\eta(D_k),
\]
where the first equality is based on Proposition~3.9~in Ref.~\cite{LLJD12-1}, 
the second one makes use of {\bf~Lemma 3.2} and the third one is due 
to the fact that $\eta(D_k)\cdot\eta(D_{k^\prime})=0$ for 
$k\neq k^\prime$. 
%By reversing the roles which $K$ and $K_0\backslash K$ play,
Moreover, since $\eta(X)=\sum_{k\in K}\eta(D_k)$, one has
$\eta(X^c)=\eta(\bigcap_{k\in K} D_k^c)=\sum_{j\in K_0\backslash K}\eta(D_j)$. Therefore, by conjugateness (Proposition~3.7~in Ref.~\cite{LLJD12-1}),
\[
\rho(X)=\neg \eta(X^c)=\neg \sum_{j\in K_0\backslash K}\eta(D_j)
 =\prod_{k\in K_0\backslash K}\neg\eta(D_k),
\] 
where 
$\neg\eta(D_k)=\rho(D_k^c)$. %$\blacksquare$
\item
Let $\forall X_l\in E_F$ $X_l=\bigcup_{k\in K^l} D_k=\bigcap_{k\in K_0\backslash K^l} G\backslash D_k$, where $K^l\subseteq K_0$.
Thus,
\begin{eqnarray}
\bigcup_i X_i=\bigcup_i\bigcup_{k\in K^i} D_k &\therefore&
\eta(\bigcup_i X_i)
%=\eta(\bigcup_i\bigcup_{k\in K^i} D_k)
=\sum_i\lrp{\sum_{k\in K^i}\eta(D_k)}=\sum_i\eta(X_i),\ 
\rho(\bigcup_i X_i)
=\sum_i\lrp{\sum_{k\in K^i}\rho(D_k)}=\sum_i\rho(X_i)
\nonumber\\
\bigcap_j X_j=\bigcap_j\bigcap_{k\in K^j} D_k^c &\therefore&
\rho(\bigcap_j X_j)
%=\rho(\bigcap_j\bigcap_{k\in K^j} G\backslash D_k)
=\prod_j\lrp{\prod_{k\in K^j}\rho(D_k^c)}=\prod_j\rho(X_j),\  \eta(\bigcap_j X_j)
=\prod_j\lrp{\prod_{k\in K^j}\eta(D_k^c)}=\prod_j\eta(X_j),\nonumber
\end{eqnarray}
where the above results as well as Proposition~3.9 of Ref.~\cite{LLJD12-1} are recursively utilised.
%$\blacksquare$
\end{itemize}
}
Another significant point is about 
the decompositon of freedom incorporated in $M^\ast$.
%Thus, in order to be sufficiently general it deserves clarifications.
If the attributes in $M$ are independent 
(not exclusive) then any attribute pairs in $M$ intersect non-trivially, giving rise to $2^{|M|}$ disjoint regions 
on the Venn diagram ${\cal V}_M^0$.
$|M^\ast|$ reaches the maximum value $2^{2^{|M|}}$ because 
any of the generalised attributes corresponds to a disjunction
of some of the $2^{|M|}$ disjoint regions on ${\cal V}_M^0$. 
In general, the members in $M$ may be intrinsically related 
such that {$r(\geq 0)$} of the disjoint regions vanish
on the Venn diagram ${\cal V}_M$, thus, 
$|M^\ast|=2^{2^{|M|}}/2^r\leq 2^{2^{|M|}}$.
Notably, disjoint regions on the Venn diagram
correspond to the $M^\ast$-atoms collected in
$b_{inf}(M^\ast)=\lbrace \mu\in M^\ast\mid \mu\succ_M {\bf 0}\rbrace$
(Lemma~2.12, Corollary~2.13~in Ref.~\cite{LLJD12-1}). 
%where $r=0$ if 
%$M$ is a collection of independent attributes.
One may henceforth adopt $e_k \in 
b_{inf}(M^\ast)$, where $e_k\cdot e_{k^\prime}\stackrel{\scriptscriptstyle k\neq k^\prime}{=}{\bf 0}$, as a convention for basis: 
\begin{eqnarray}
e_k\equiv e_k(M),&& k\in K^\ast:=\lbrace 1, 2, \ldots, rank(M)\rbrace, 
\label{eq:e_k}\\
rank(M)&=&|b_{inf}(M^\ast)|=\log_2^{|M^\ast|}=2^{|M|}-r.\nonumber
\end{eqnarray}
%for $k\in K^\ast:=\lbrace 1, 2, \ldots, \log_2^{|M^\ast|}\rbrace$
%where each $\prod \Psi_M^k$ denotes a non-trivial infimum for $M^\ast$
%(Definition 2.6, Lemma 2.12 \cite{LLJD12-1})
%as it is impossible to find $\mu\in M^\ast$ s.t. $0<\mu<\prod \Psi_M^k$. 
%where $e_k \in b_{inf}(M^\ast)$
%where %$rank(M)$ %$:=|b_{inf}(M^\ast)|=\log_2^{|M^\ast|}=2^{|M|}-r$
%is referred to as 
%the {\it rank} of the generalised attribute set $M^\ast$.
%This is because 
Note that $rank(M)$ is a particular integer 
describing the freedom in $M^\ast$ subject to the
{\it intrinsic constraint} held among members in $M$
and $e_k\equiv e_k(M)$ is to mark that $e_k$ is  
also constructed out of $M$. 
%(Definition~2.7~in Ref.~\cite{LLJD12-1}).
However, it turns out formally
\[
M^\ast=(b_{inf}(M^\ast))^\ast=(M^\ast)^\ast,\quad rank(M)=rank\left(b_{inf}(M^\ast)\right)=rank(M^\ast),%\quad e_k(M)=e_k(b_{inf}(M^\ast))=e_k(M^\ast), 
\]
highlighting that the set $M$ which
gives rise to the generalised attribute set $M^\ast$ 
with $rank(M^\ast)=2^{|M|}-r$ need not be unique.
Indeed, there are abundant choices of $M^\ast$
fulfilling such a requirement, 
which will furnish the framework 
for analysing the reparametisation of the formal context 
in the next paper.   
Here, an immediate example appears to be  
the choice $M_0\equiv b_{inf}(M^\ast)$ such that 
$M_0^\ast=M^\ast$ whenever $M\neq b_{inf}(M^\ast)$.
Certainly, $M_0$ then manifests a {\it intrinsic constraint} 
in the sense that it only consists mutually exclusive 
members.
%$M_0^\ast$ as well gives the full attribute set 
%constructable out of $M$. 
%This is such as that $M^\ast$ is the full attribute set 
%constructable out of $M$.
%Notably, there are $2^{|M|}$ $M^\ast$-atoms for the generalised attribute-set $M^\ast$
%(Lemma 3.10 \cite{LLJD12-1}). 

Turning back to the lattice construction, 
a primary concern is about how the {\bf GCL}
presents itself.
\dfn{
%{\bf Definition~3.4}\\*
The {\bf GCL}
subject to a formal context will be henceforth referred to as 
$\Gamma_F(G,M)$, cf. Poposition~3.16~in Ref.~\cite{LLJD12-1}, which is uniquely prescribed by either of 
\begin{itemize}
\item
the $\eta$-representation 
${\Upsilon}_\eta^F(G,M):= [\eta(D_1),\ldots,\eta(D_{n_F})]\equiv [\eta_1,\ldots,\eta_{n_F}]$,
\item
the $\rho$-representation ${\Upsilon}_\rho^F(G,M):=[\rho(G\backslash D_1),\ldots,\rho(G\backslash D_{n_F})]\equiv [\neg \eta_1,\ldots,\neg \eta_{n_F}]$,
\end{itemize}
where $n_F$ denotes the number of subclasses discernible from 
the point of view of $F(G,M)$.} 
%while $0_\rho=\prod_k{\hat 0}_\rho^k$ and $1_\eta=\sum_k {\hat 1}_\eta^k$, 
%where ${\hat 0}_\rho^k ({\hat 1}_\eta^k)$ is the $k$th component of ${\hat 0}_\rho ({\hat 1}_\eta)$.
%${\hat 0}_\rho $ and ${\hat 1}_\eta$
%thus overwhelm the roles which ${\cal P}_\eta$ and ${\cal P}_\rho$ originally play.
%from which one realises that 
%Moreover, ${\hat 0}_\rho$ can be determined from ${\hat 1}_\eta$ (or vice versa) by virtue of ${\hat 0}_\rho^k=\neg {\hat 1}_\eta^k$ since $\rho(G\backslash D_k)=\neg\eta(D_k)$.
%Call, henceforth, ${\hat 1}_\eta$ 
%the "$\eta$-represention" and 
%${\hat 0}_\rho $ the "$\rho$-represention" for the {\bf GCL}.  
According to {\bf Proposition~3.1}, obtaining the $\eta$-representation or the $\rho$-representation is a simple one-scan task, on the formal context.
%{\bf Proposition~3.3} then suggests Eq.~(\ref{eq:BX_k})
%that can be employed 
%to read off $\rho(X)$ and $\eta(X)$ for all 
%the general extent $X$.   
%In Fig. \ref{fig:Calculate_GfcpGrsp}
For the $F(G,M)$ in Table \ref{table:formal_scheme}, 
it is straightforward to write down 
\[
{\Upsilon}_\eta^F\ =[\underbrace{a\neg bcde}_{\eta(D_1)},\underbrace{a\neg bc\neg d\neg e}_{\eta(D_2)},
\underbrace{\neg ab\neg c\neg de}_{\begin{smallmatrix}
\eta(D_3)\\
D_3=\lbrace 3,4\rbrace
\end{smallmatrix}},\underbrace{a\neg b\neg c\neg d\neg e}_{\eta(D_4)},
\underbrace{ab\neg c\neg de}_{\eta(D_5)}].
\]
%$
%{\bf 1}=\sum_{k=1}^{\log_2^{|M^\ast|}} e_k.
%$
Note that
{\bf Proposition~3.3} in fact provides simple 
identifications for the whole {\bf GCL} structure.
For convenience,
assume that it is the first $n_F$ $M^\ast$-atoms 
in the convention of Eq.~(\ref{eq:e_k})
which enter $\eta(X)$ as constituents: 
\begin{equation}
\eta(D_k)\equiv \eta_k=\prod \Psi^k:=e_k,\ k\in K_0=\lbrace 1, 2, \ldots, n_F\rbrace\subseteq K^\ast, \label{eq:etaDk}
\end{equation}
where $\Psi^k=\lbrace m\in M \mid m\in D_k^I\rbrace\cup 
\lbrace \neg m \mid m\not\in D_k^I, m\in M\rbrace$ (cf. Eq.~(\ref{eq:etadnf})).
Then,
%Intuitively, each allowable extent for {\bf GCL} unambiguously corresponds to a binary string.
$\eta(X)=\eta(\bigcup_{k\in K} D_k)=\sum_{k\in K}\eta_k$, which
picks up a subset $K\subseteq K_0$ from the expression $1_\eta=\sum_{k\in K_0}\eta_k$ ({\bf Proposition 3.1}).
Likewise, 
%self-duality asserts that 
$\rho(\bigcap_{k\in K_0\backslash K} G\backslash D_k)
=\prod_{k\in K_0\backslash K}\neg \eta_k$
picks up all the components which are {\it not} in $K$ from the expression 
$0_\rho=\prod_{k\in K_0} \neg \eta_k$
since $\rho(G\backslash D_k)=\neg \eta_k\ \forall k \in K$.
%Remarkably, $X=\bigcup_{k\in K} D_k\equiv \bigcap_{k\in K_0\backslash K} G\backslash D_k$ simultaneously 
%allocates the attribute-units of concern for both $\eta(X)$ and $\rho(X)$ since mutually complementary 
%index-subsets are employed.
It is then straightforward to extract the general extents and intents 
via $n_F$-bit binary masks from a known {\bf GCL} structure 
as follows.
Let $B_X$ be the binary string whose $k$th bit $B_X^k$ is given as
$
\left\lbrace
\begin{smallmatrix}
B_X^k=1\ \mbox{if}\ D_k\subseteq X\\
B_X^k=0\ \mbox{if}\ D_k\not\subseteq X
\end{smallmatrix}\right.$.
Then, for any $X\in E_F$ there is a binary string 
$B_X$ marking the $D_k$'s that $X$ contains;  
any of $n_F$-bit binary strings corresponds to a 
definite general extent.
% since $E_F$ consists of all possible unions 
%of $D_k$s as well as the empty set $\emptyset$. 
%The triplet $(X,\rho(X),\eta(X))$ is thus identified as
One may thus write down that
\begin{eqnarray}  
X &=&\bigcup \lbrace D_k \mid B_X^k=1,\ k\in K_0  \rbrace,\nonumber\\
\eta(X)&=&[B_X]_\eta=\sum \lbrace \eta_k \mid B_X^k=1,\ k\in K_0  \rbrace,\nonumber\\ 
\rho(X)&=&[B_X]_\rho=\prod \lbrace \neg \eta_k \mid B_X^k=0.\ k\in K_0  \rbrace.
\label{eq:BX_k}
\end{eqnarray}
%In Fig. \ref{fig:Calculate_GfcpGrsp}.
Thus,  one has $B_G=$\q{11111} which implies
that $\eta(G)=[11111]_\eta=1_\eta$ in the above example, where 
$1_\eta=\sum {\Upsilon}_\eta^F$
sums up all the components of the $\eta$-representation of {\bf GCL} ({\bf Definition~3.4}).  
Likewise, $B_\emptyset$=\q{00000} tells that 
$\rho(\emptyset)=[00000]_\rho=0_\rho$ where 
$0_\rho=\neg 1_\eta=\prod {\Upsilon}_\rho^F$ is the product of all the components of the $\rho$-representation.
Let one also consider that $\rho(X)=\prod_k\neg e_k\equiv\prod_k\neg e_k\cdot {\bf 1}$ and $\eta(X)=\sum_k e_k\equiv\sum_k e_k+{\bf 0}$.
In such a manner, $\rho(G)=[11111]_\rho=\bf{1}$ 
because one picks up no
term in $\lbrace \neg e_1,\ldots,\neg e_5\rbrace$ due 
to the absence of {0} in $B_G$.
Similarly, $\eta(\emptyset)=[00000]_\eta=\bf{0}$ due to the absence of {1} in $B_\emptyset$.
%as no "1" is found in $\lbrace e_1,\ldots,e_5\rbrace$.
%The results for $\rho(G)$ and $\eta(\emptyset)$ can be made intuitive:
%The above construction in fact means that the {\bf GCL} is completely determined up to 
%the following fundamental quantities.\\

\begin{figure}   
\includegraphics[scale=0.6,angle=-90]{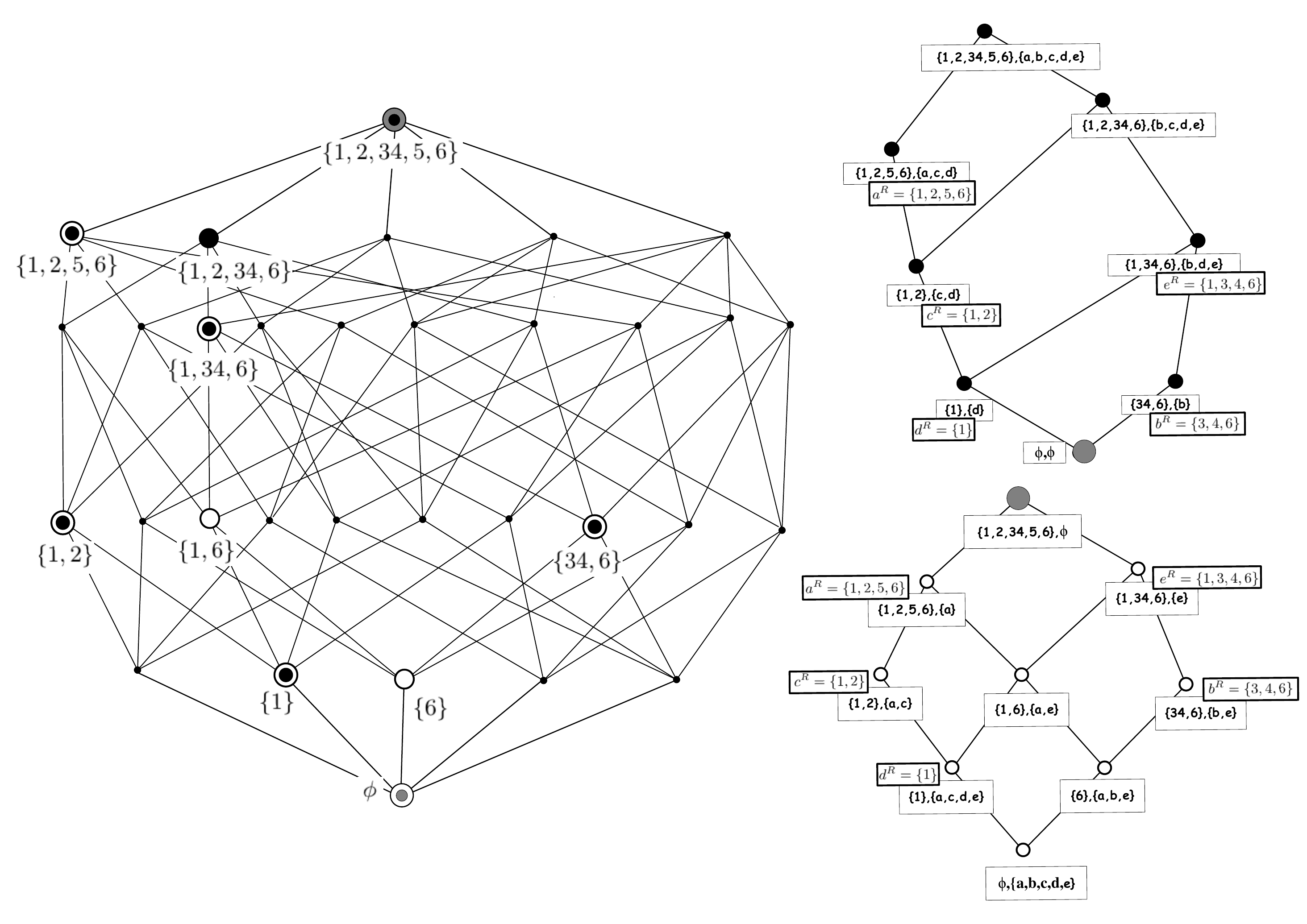}
\caption{
Rediscovering the traditional concept lattices on the {\bf GCL}\\
The circled points are nodes existing on the FCL, whereas the bold ones belong to the RSL.  
To determine their intents, note that
the FCL-intent can always be found from 
the lower bound in {\bf CNF} of the corrsponding general intent
and RSL-intent the upper bound the {\bf DNF}, see Proposition~3.15 
in Ref.\cite{LLJD12-1}.
Consider, e.g.,
${[11011]_\eta}=
  {\bf a}\neg bcde+{\bf a}\neg bc\neg d\neg e+{\bf a}\neg b\neg c\neg d\neg e+{\bf a}b\neg c\neg de
\equiv %{\bf a}(b+c)(d+\neg e)(\neg c+d)(b+\neg e)(\neg b+\neg d+e)
{\bf a}(\neg b+e)(\neg d+e)
(\neg b+\neg c)(\neg b+\neg d)(c+\neg d)
(b+c+\neg e)
(b+d+\neg e)
(\neg c+d+\neg e)$ and
${[11011]_\rho}={\bf a}+\neg b+{\bf c}+{\bf d}+\neg e$, as can be determined in the manner of Fig~\ref{fig:Calculate_GfcpGrsp}.
Now that ${\bf a}^R=\lbrace a \rbrace^I=({\bf a+c+d})^R=\lbrace {a,c,d} \rbrace^\Diamond=\lbrace 1,2,5,6 \rbrace$, one rediscovers the formal concepts 
$(\lbrace 1,2,5,6 \rbrace,\lbrace a\rbrace)$ for FCL 
and $(\lbrace 1,2,5,6 \rbrace,\lbrace a,c,d\rbrace)$ for RSL.
Clearly, it is unwarranted that both the FCL and RSL
possess common extents,
the collection of common extents for RSL and FCL essentially includes 
$\lrbr{a^R,b^R,c^R,d^R,e^R}$ (see Proposition~3.2~in~Ref.~\cite{LLJD12-1}).
Take for instance the object set $\lbrace 1,6 \rbrace$, which 
is an FCL-extent but {\it not} an RSL-extent.
Here, $\eta(\lbrace 1,6 \rbrace)=a\neg bcde+ab\neg c\neg de=
ae(\neg bcd+b\neg c\neg d)$ leads to $\lbrace 1,6 \rbrace^I=\lrbr{a,e}$
since $(ae)^R=\lrbr{a,e}^I=\lrbr{1,6}$.
On the other hand,
$\rho(\lbrace 1,6 \rbrace)=(\neg a+b+\neg c+{d}+e)(a+\neg b+c+{d}+\neg e)(\neg a+b+c+{d}+e)\equiv {d}+ab+ce+\neg be+\neg a\neg e$
seems to suggest the concept 
$(\lbrace 1,6 \rbrace,\lbrace d \rbrace)$ for RSL.
However, it fails, as
$d^R\equiv \lbrace d\rbrace^\Diamond= 
\lbrace 1\rbrace\neq \lbrace 1,6\rbrace$.}
\label{fig:rslfcl_gcl}
\end{figure}

It is also particularly interesting to rediscover the FCL and RSL within a known {\bf GCL} structure (Fig. \ref{fig:rslfcl_gcl}).
%Starting with a known {\bf GCL} structure, 
%two steps are considered according to 
%which
%then automatically gives rise to 
%$\eta (X)=\prod_{X_0\supseteq X}\left(\prod [X_0^{+}]_F\right)$ and
%$\rho(X) =\sum_{X_0\subseteq X}\left(\sum [X_0^{\times}]_F\right)$.
After Proposition~3.15~in Ref.~\cite{LLJD12-1},
two steps are in order.
Firstly, re-express
the given $\eta(X)\ \mbox{in {\bf CNF}}$ and $\rho(X)\ \mbox{in {\bf DNF}}$.
Then, one may 
collect from 
these expressions 
%all the $[X_0^{+}]_F$'s~({$[X_0^{\times}]_F$'s}) of concern in $\eta(X)$~({$\rho(X)$})
those attributes 
belonging to $M$
%to collect 
to {form} a {\it candidate} FCL intent $Y_{c-fcl}(X)$
and a candidate RSL intent $Y_{c-rsl}(X)$, respectively.
Secondly, if $(Y_{c-fcl}(X))^I=X$ then 
$(X, Y_{c-fcl}(X))$ is accepted as an FCL concept, whereas
if $(Y_{c-rsl}(X))^{\Diamond}=X$ then
$(X, Y_{c-rsl}(X))$ is accepted as an RSL concept.
%say $(X,Y(X))_{fcl}$~($(X,Y(X))_{rsl}$).
%Note that the condition $(Y_{c-fcl}(X))^I=X$~($(Y_{c-rsl}(X))^{\Diamond}=X$)
%can be operated by means of $\bigcap_{m\in Y_{c-fcl}(X)} m^R=X$~($\bigcup_{m\in Y_{c-rsl}(X)} m^R=X$) (Proposition~3.15 \cite{LLJD12-1}).
Note that such constructions could be less intuitive for particular nodes.
%since the application of the conditions
%\q{$\left\lbrace \begin{smallmatrix}
 %X^I=Y\\
 %Y^I=X 
 %\end{smallmatrix}\right.$,%(for the FCL-concept)
 %$\left\lbrace \begin{smallmatrix}
 %X^\Box=Y\\
 %Y^\Diamond=X 
 %\end{smallmatrix}\right.$} %(for the RSL-concept)
%based on Eq. (\ref{eq:derivation_O}) 
%are {\it not} always well defined for any ($X$, $Y$).
% is an empty set.
For instance, 
%since $\eta (\emptyset)={\bf 0}$, 
at $X=\emptyset$,
one should imagine 
$\eta (\emptyset)={\bf 0}=a\cdot b\cdot c\cdot d\cdot e\cdot \ldots$
(Step 1) so one  
ends up with the FCL concept $(\emptyset,\lbrace a,b,c,d,e\rbrace)$, based on 
$a^R\cap b^R \cap c^R\cap d^R\cap e^R=\emptyset$ (Step 2).
Similarly, $\rho(G)={\bf 1}=a+b+c+d+e
+\ldots$ and
$a^R\cup b^R \cup c^R\cup d^R\cup e^R=\lbrace 1,2,3,4,5,6\rbrace=G$ imply that
$(\lbrace 1,2,3,4,5,6\rbrace,\lbrace a,b,c,d,e\rbrace)$ is a RSL concept.
It seems
by Proposition~3.15 of Ref.~\cite{LLJD12-1}
that 
\[
E_{F}^{fcl}=\lrbr{\bigcap_{m\in M_0}m^R \mid M_0\subseteq M},\
E_{F}^{rsl}=\lrbr{\bigcup_{m\in M_0}m^R \mid M_0\subseteq M}
\]
provide all the FCL and RSL extents.
However, conventionally, 
one also regards the object class $G$ as an FCL extent
and $\emptyset$ as an RSL extent, while, as object classes, 
$G\not\in E_{F}^{fcl}$ and $\emptyset\not\in E_{F}^{rsl}$.
The point is that  
the FCL concept corresponding to $G$
do not always satisfy 
the condition $\left\lbrace \begin{smallmatrix}
X^I=Y\\
Y^I=X 
\end{smallmatrix}\right.$ and 
the RSL concept corresponding to $\emptyset$
do not always satisfy 
$\left\lbrace \begin{smallmatrix}
X^\Box=Y\\
Y^\Diamond=X 
\end{smallmatrix}\right.$.
%Table \ref{table:formal_scheme}
%but needs not to come as a surprise
In effect, e.g. in Fig. \ref{fig:rslfcl_gcl}, $(G,\emptyset)$ is determined via
$(\bigcup_{X\in E_{F}^{fcl}} X, \bigcap_{X\in E_{F}^{fcl}} Y(X))$
and $(\emptyset,\emptyset)$ is via $(\bigcap_{X\in E_{F}^{rsl}} X, \bigcap_{X\in E_{F}^{rsl}} Y(X))$, 
which are {\it artificially} appended as the 
lattice supremum and infimum respectively, for the sake of 
the completeness of lattices \cite{Wi82,YY04}. 
It should however be clear 
%that after the extension \q{$F(G,M)$ to $F^\ast(G,M^\ast)$}
that there is no need of such {\it artificial completions}
for the {gFCL} and {gRSL} ({\bf Proposition~2.2}).
The conditions $\left\lbrace \begin{smallmatrix}
X^{I^\ast}=Y\\
Y^{I^\ast}=X 
\end{smallmatrix}\right.$ and $\left\lbrace \begin{smallmatrix}
X^{\Box^\ast}=Y\\
Y^{\Diamond^\ast}=X 
\end{smallmatrix}\right.$
%as well as $\left\lbrace \begin{smallmatrix}
%X^{\Diamond^\ast}=Y\\
%Y^{\Box^\ast}=X 
%\end{smallmatrix}\right.$ 
are well defined because 
these conditions are all based on 
the general intents
which 
are never empty sets.
Indeed, the cardinality of general intent 
even remains constant over $E_F$, as will be demonstrated
in {\bf Corollary~4.4}.
%Remarkably, 
%the general concepts is in fact {\it free from} the consideration 
%based on with generalising the conventional FCL and RSL.
%the general concepts $(X, [X]_F)$ 
%({\bf Proposition 2.1}). 
%In particular, $[\emptyset]_F$
%can be identified as the all the attributes 
%which lie between ${\bf 0}$ and $0_\rho$ (also cf. {\bf Corollary 4.3}), while 
%$\mu^R=\emptyset$ (namely enclosing no region on ${\cal V}_M^F$) for ${\bf 0}\leq\mu\leq 0_\rho$.
%On the other hand, all the intents for {\bf GCL} are non-empty

%Note with $\eta_k=e_k$ (Eq. (\ref{eq:etaDk}))
%one has also emphasised that subject to a formal context the {\bf GCL} always ends %up with 
%$\eta_k\succ_M {\bf 0}\ \forall k\in K_0$  (Lemma 3.10 \cite{LLJD12-1}). \\ 
%Moreover, if $\mu\leq 1_\eta$ then $\mu=\sum_{k\in K}e_k$, for some $K\subseteq K_0$, essentially emerging as a part of the 
%sum $\sum_{k=1}^{n_F}e_k$.
%Hence, $|\lbrace \mu\in M^\ast\mid \mu\leq 1_\eta \rbrace|=2^{n_F}$ since this amounts to
%all possible disjunctions of the members in $\lbrace {\bf 0},e_1,\ldots e_{n_F}\rbrace$.
%Likewise, one also has $|\lbrace \mu\in M^\ast\mid \mu\geq 0_\rho \rbrace|=2^{n_F}$.\\ 

\section{implication relations}\label{four}
%Alternatively, one may consider determining extents through attribute-matching.\
It is the object-attribute relationship 
resulted from the categorisation 
that leads to 
the {\it logic significance} implemented by the {\bf GCL} structure.  
Roughly speaking, 
one's {\it inspections of attributes} are essentially restricted in a definite object domain thus the attributes
receive additional ordering prescriptions, which are the origins 
of the logic implication in {\bf GCL}.
Here, as a primary observation, the attributes that play the roles 
of the bounds of general intents are equipped with particular features:
%Given a {\bf GCL} subject to the formal context $F(G,M)$,
\begin{equation}
\forall \mu\in M^\ast\quad
\begin{array}{cc}
\mu\leq 1_\eta \iff \mu=\eta (X)& \mbox{for some}\ X\in E_F \\
\mu\geq 0_\rho \iff \mu=\rho (X)& \mbox{for some}\ X\in E_F
\end{array},\label{eq:mu_upper_lower}
\end{equation}
which is a direct consequence of {\bf Proposition~3.3}.
In what follows, 
by matching any property of a given object with a {\bf GCL} one
can determine its class belongingness in the {\bf GCL} structure.
%If $\mu=\eta (X)$ then it can be written as some $[B_X]_\eta$ 
%which is smaller than $1_\eta=[1\ldots1]_\eta$, where \q{$1$} has been repeated $n_F$ times.
%Conversely, if $\mu\leq 1_\eta$ then $\mu=\sum_{k\in K}e_k$
%for some $K\subseteq K_0$ which means
%$\mu=[B_X]_\eta$ for some $B_X$ corresponding to certain {\it GE}, say $X$. 
%By self-duality, $\forall \nu$ which is a {\it Gfcp}, $\neg \nu$ is a {\it Grsp} and vice versa.
%Consequently, $\nu\leq 1_\eta$ or $\neg \nu\geq \neg 1_\eta\equiv 0_\rho$ 
%iff $\neg \nu$ is a {\it Grsp}. Put $\mu=\neg \nu$. 
%{\bf Proposition~4.1}\\
\prp{
Given a {\bf GCL} subject to the formal context $F(G,M)$,  
$\forall \mu\in M^\ast\ (\mu^R,[\mu\cdot 1_\eta,\mu+0_\rho])$ 
is a general concept, cf. {\bf Proposition~2.1} and 
Eq.~(\ref{eq:g_concept_23}),
every general concept of the {\bf GCL} can 
be unambiguously represented by $(\mu^R,[\mu^R]_F)$ 
with some $\mu\in M^\ast$.}
%{\bf Proof}: \\
%It is obvious that $\mu+0_\rho$ is a {\it Grsp} and $\mu\cdot 1_\eta$ is a {\it Gfcp}.
\prf{
Equation~(\ref{eq:mu_upper_lower}) implies that
$\forall\mu$, $\mu\cdot 1_\eta$ is a lower bound of some general intent, i.e. $\eta(X)$ with some $X\in E_F$,
because $\mu\cdot 1_\eta\leq 1_\eta$. 
%$[B]_\eta$ for some binary string $B$ of $n_F$ bits (Eq. \ref{eq7}).
Likewise, $\mu+ 0_\rho$ can be identified as the upper bound 
$\rho(X^\prime)$ with some $X^\prime\in E_F$.
% $[B^\prime]_\rho$ for 
%some binary string $B^\prime$ of $n_F$ bits.
%Because each $n_F$-bit binary string represents a {\it GE}, 
%one may identify $B\equiv B_X$ and $B^\prime\equiv B_{X^\prime}$ for some $X$ and $X^\prime$. 
Moreover, $X^\prime=(\mu+0_\rho)^R=\mu^R=(\mu\cdot 1_\eta)^R=X$.
Therefore, 
$(\mu\cdot 1_\eta)=\eta(\mu^R)$ and $(\mu+0_\rho)=\rho(\mu^R)$ implies that 
$(\mu^R,[\mu\cdot 1_\eta,\mu+0_\rho])$ is always a general concept.
%since $\mu^R\in E_F$.
On the other hand, 
%Subsequently, by given $(X,\rho(X),\eta(X))$,
%$\mu$ that renders $(\mu^R,\mu+0_\rho,\mu\cdot 1_\eta)=(X,\rho(X),\eta(X))$ a general context indeed exists.
%If {\bf S1} is true, 
for any given $X\in E_F$, 
any $\mu \in [X]_F$ will render $(\mu^R,[\mu^R]_F)$ a general concept, as is demonstrated above.
This is not ambiguous since 
general intents are disjoint, namely, 
$\forall X_i\forall X_j \in E_F\ [X_i]_F\cap [X_j]_F=\emptyset\iff X_i\neq X_j$  ({\bf Proposition~2.1}).} %%$\blacksquare$\\
%Note such construction is not ambiguous 
%since $\rho(X)$ and $\eta(X)$ are unique with respect to $X$ (Proposition 3.6 \cite{LLJD12-1}).\\
%then $\mu+0_\rho\geq \nu\geq \mu\cdot 1_\eta\ \forall \nu \in [X]_F$.
%Namely, $\mu+0_\rho$ and $\mu\cdot 1_\eta$ are respectively the upper and the lower bound $[X]_F$ and 
%thus can be identified as $\rho(X)$ and $\eta(X)$. ({\bf Proposition 2.1}).
%Therefore, {\bf S1} $\Rightarrow$ {\bf S2}.
%On the other hand, assume it is not true that {\bf S2} $\Rightarrow$ {\bf S1}: 
%Every $(X,\rho(X),\eta(X))$ can be written as $(\mu^R,\mu+0_\rho,\mu\cdot 1_\eta)$ 
%but $\exists \mu_0\in M^\ast $ s.t. $(\mu_0^R,\mu_0+0_\rho,\mu_0\cdot 1_\eta )$ 
%is {\it not} a general concept for {\bf GCL}.
%Then, $\mu_0^R$ can be identified as some $X_0$ and
%$\mu_0+0_\rho\neq \rho(X)$ and/or $\mu_0\cdot 1_\eta\neq \eta(X)$. 
%Contradiction arises in the following manner. 
%While $\mu_0+0_\rho=[{X^\prime}]_\rho$ and 
%$\mu_0\cdot 1_\eta=[B_{X^{\prime\prime}}]_\rho$, 
%not both $X^\prime$ and $X^{\prime\prime}$
%can be identified as $X_0$, however, $(\mu_0+0_\rho)^R=(\mu_0\cdot 1_\eta)^R=X_0$.
%\end{itemize}
%Certainly, obtaining {\it Grsp} and {\it Gfcp} through $\mu$ meets abundant choices.\\
%{\bf Corollary 4.2}\\*
\crl{
Given a formal context $F(G,M)$, $\forall X \in E_F$,
\begin{itemize}
\item $\rho(X)=\eta(X)+0_\rho$ and $\eta(X)=\rho(X)\cdot 1_\eta$,
\item $\rho(X)\cdot \neg \eta(X)=0_\rho$ and $\neg \rho(X)+\eta(X)=1_\eta$.
\end{itemize}
}
\prf{
%{\bf Proof}:
\begin{itemize}
\item 
According to {\bf Proposition~4.1}, substituting $\mu=\eta(X)$ 
and $\mu=\rho(X)$ 
into the expression 
$(\mu^R, [\mu\cdot 1_\eta,\mu+0_\rho])$ 
must lead to the same general concept 
since $\rho(X)^R=\eta(X)^R=X$.
Consequently, 
$(X,[\eta(X),\eta(X)+0_\rho])=(X,[\rho(X)\cdot 1_\eta,\rho(X)])$
implies that $\eta(X)+0_\rho=\rho(X)$ and $\eta(X)=\rho(X)\cdot 1_\eta$.
%$\blacksquare$
\item $\eta (X)$ and $\rho (X)$ can be respectively identified as $ \mu\cdot 1_\eta$ and $\mu+0_\rho$  for some $\mu\in M^\ast$.
Therefore, $\rho(X)\cdot \neg \eta(X)=(\mu+0_\rho)\cdot \neg (\mu\cdot 1_\eta)=(\mu+0_\rho)\cdot (\neg \mu+ 0_\rho)=0_\rho$.
On the other hand, $\neg \rho(X)+\eta(X)= \neg(\rho(X)\cdot \neg \eta(X))=1_\eta$. %$\blacksquare$  
\end{itemize}
}
Thus, $0_\rho$ and $1_\eta$ can furnish the connection between the upper and lower bounds for any $[X]_F$.  
In addition, $\rho(X)\cdot \neg \eta(X)$ and $\neg \rho(X)+\eta(X)$ are in fact {\it constants} over the full {\bf GCL}.
%{\bf Corollary 4.3}\\
\crl{
The general intent $[X]_F$ is the equivalent class of attributes 
generated from the pair ($\rho(X),\eta(X)$):
%Subject to the formal context $F(G,M)$,
\[
\forall X\in E_F\quad
[X]_F=\lbrace \mu\in M^\ast\mid \mu+0_\rho=\rho(X),\ \mu\cdot 1_\eta=\eta(X)\rbrace.
\]
}
%{\bf Proof}:\\
\prf{
For all $\mu_1\in\lbrace \mu\in M^\ast\mid \mu+0_\rho=\rho(X),\ \mu\cdot 1_\eta=\eta(X)\rbrace$ 
one has the general concept 
$(\mu_1^R,\rho(\mu_1^R),\eta(\mu_1^R))$ in which $\mu_1^R=X$, i.e. $\mu_1\in [X]_F$ ({\bf Proposition~4.1}).
Thus, $\mu_1\in\lbrace \mu\in M^\ast\mid \mu+0_\rho=\rho(X),\ \mu\cdot 1_\eta=\eta(X)\rbrace\implies \mu_1\in [X]_F$.
On the other hand,
if
$\mu_1\in\lbrace \mu\in M^\ast\mid \mu+0_\rho=\rho(X), \mu\cdot 1_\eta=\eta(X)\rbrace$ then $\mu_1^R=X$.
\[
\mu_1 \not \in [X]_F\implies \mu_1^R\neq X\ \lrp{\because\lsmatr{(\mu_1\cdot 1_\eta)^R=\mu_1^R\cap G=\mu_1^R\neq X\\
(\mu_1+0_\rho)^R=\mu_1^R\cup \emptyset\neq X}}\ \mbox{which implies}\
\mu_1\not\in\lbrace \mu\in M^\ast\mid \mu+0_\rho=\rho(X),\ \mu\cdot 1_\eta=\eta(X)\rbrace.
\]
%If $\mu_1 \not \in [X]_F$ 
%then $\mu_1^R\neq X$, i.e. 
%$\lsmatr{(\mu_1\cdot 1_\eta)^R=\mu_1^R\cap G=\mu_1^R\neq X\\
%\mu_1+0_\rho)^R=\mu_1^R\cup \emptyset\neq X}$, $\therefore \mu_1\not\in\lbrace \mu\in M^\ast\mid \mu+0_\rho=\rho(X),\ \mu\cdot 1_\eta=\eta(X)\rbrace$.
Therefore, $\forall \mu_1\in M^\ast,\ \mu_1 \in [X]_F\iff\mu_1\in\lbrace \mu\in M^\ast\mid \mu+0_\rho=\rho(X),\ \mu\cdot 1_\eta=\eta(X)\rbrace$.} %%$\blacksquare$\\
It is also notworthy that 
by the $n_F$ general intents one in fact exhausts 
the whole generalised attribute set $M^\ast$
in view of $2^{(\log_2^{|M^\ast|}-n_F)}\times 2^{n_F}=|M^\ast|$.
\crl{
%{\bf Corollary 4.4}
Given a collection of general intents, say $\lrbr{[X_1]_F,[X_2]_F,\ldots [X_n]_F}$,
\begin{itemize}
\item
if $\mu_i\in [X_i]_F$ for $1\leq i\leq n$ then
$\sum_{i=1}^n \mu_i\in \lrsqp{\bigcup_{i=1}^n X_i}_F$ as well as 
$\prod_{i=1}^n \mu_i\in \lrsqp{\bigcap_{i=1}^n X_i}_F$,
\item
all the general intents for the general concept lattice 
are of the same cardinality:
$|[X]_F|=2^{(\log_2^{|M^\ast|}-n_F)}\ \forall X \in E_F$ subject to the formal context $F(G,M)$.
\end{itemize}
}
%{\bf Proof}: 
\prf{
\begin{itemize}
\item 
Since $\mu_i\in [X_i]_F$, one has $\mu_i=\eta(X_i)+\tau_i$ with some
$\tau_i\in [{\bf 0}, 0_\rho]$ by {\bf Proposition~2.1}.  
Therefore, 
\[
\begin{array}{ccccc}
\sum_{i=1}^n \mu_i&=&\sum_{i=1}^n\eta(X_i)+\sum_{i=1}^n\tau_i=
\eta\lrp{\bigcup_{i=1}^n X_i}+\tau&\in& \lrsqp{\bigcup_{i=1}^n X_i}_F,\\
\prod_{i=1}^n \mu_i&=&\prod_{i=1}^n\eta(X_i)+\prod_{i=1}^n\tau_i=
\eta\lrp{\bigcap_{i=1}^n X_i}+\tau^\prime&\in& \lrsqp{\bigcap_{i=1}^n X_i}_F
\end{array}
\]
%\begin{eqnarray}
%\sum_{i=1}^n \mu_i=\sum_{i=1}^n\eta(X_i)+\sum_{i=1}^n\tau_i=
%\eta\lrp{\bigcup_{i=1}^n X_i}+\tau&\in& \lrsqp{\bigcup_{i=1}^n X_i}_F,
%\nonumber\\
%\prod_{i=1}^n \mu_i=\prod_{i=1}^n\eta(X_i)+\prod_{i=1}^n\tau_i=
%\eta\lrp{\bigcap_{i=1}^n X_i}+\tau^\prime&\in& \lrsqp{\bigcap_{i=1}^n X_i}_F
%\nonumber
%\end{eqnarray}
by {\bf Proposition~3.3}
since
$\tau=\sum_{i=1}^n\tau_i\in [{\bf 0}, 0_\rho]$
and $\tau^\prime=\prod_{i=1}^n\tau_i\in [{\bf 0}, 0_\rho]$. %$\blacksquare$
\item
$\rho(X)=\eta(X)+0_\rho\ \forall X \in E_F$ ({\bf Corollary 4.2}). 
Hence, any $\mu \in [X]_F\equiv [\eta(X),\eta(X)+0_\rho]$ can be written as $\mu=\eta(X)+\tau$ 
with ${\bf 0}\leq\tau\leq 0_\rho$.
Upon employing the convention of Eq. (\ref{eq:e_k}) and (\ref{eq:etaDk}), one may identify 
$0_\rho=\sum_{k=n_F+1}^{\log_2^{|M^\ast|}}e_k$ 
since $0_\rho\cdot 1_\eta={\bf 0}$ and $0_\rho+ 1_\eta={\bf 1}$. 
Subsequently, any $\tau\in [{\bf 0},0_\rho]$ can be written as $\tau=\sum_{k\in Q}e_k$ 
for some $Q\subseteq K^\ast\backslash K_0=\lbrace n_F+1,\ldots,\log_2^{|M^\ast|}\rbrace$.
Therefore, $|[X]_F|=|[{\bf 0},0_\rho]|=|2^{K^\ast\backslash K_0}|=2^{(\log_2^{|M^\ast|}-n_F)}$. %$\blacksquare$
\end{itemize}
}
According to {\bf Proposition 4.1},
%The reason why $(\mu+0_\rho)^R$ and $(\mu\cdot 1_\eta)^R$ coincide is 
%interpreted as follows.
%$(\mu+0_\rho)^R$ is the {\it smallest} object-set $X$ 
%such that $\mu$ is not found in $X^c$,
%while $(\mu\cdot 1_\eta)^R$ is the {\it largest} object-set that possesses the feature $\mu$ in common.
an object set $X=\mu^R$ for which $\mu\in [X]_F$ 
can be 
acquired by   
$\mu+0_\rho=[B_X]_\rho$ or $\mu\cdot 1_\eta=[B_X]_\eta$, see Eq.~(\ref{eq:BX_k}).
For instance, consider $\mu=a\cdot \neg e+c$ in  
Fig.~\ref{fig:rslfcl_gcl}. Then,
\begin{eqnarray}
(a\cdot \neg e+c)\cdot 1_\eta&=&a\neg bcde+a\neg bc\neg d\neg e+a\neg b\neg c\neg d\neg e=[11010]_\eta,\nonumber\\
(a\cdot \neg e+c)+ 0_\rho&=&(a+\neg b+c+d+\neg e)(\neg a+\neg b+c+d+\neg e)=[11010]_\rho \nonumber
\end{eqnarray}
both imply $B_X=$\q{11010}, hence $X=\lbrace 1,2,5\rbrace$, which could also be deduced from $\mu^R=a^R\cap e^{Rc}\cup c^R=\lbrace 1,2,5\rbrace$.
%Let one also summarise the fundamental characteristics for {\bf GCL} in terms 
%of function-dependencies subject to $F(G,M)$, say,
%\begin{eqnarray}  
%\Gamma_F&\equiv& \Gamma_F({\hat 1}_\eta)\ \mbox{or}\ \Gamma_F\equiv \Gamma_F({\hat 0}_\rho),\nonumber\\*
%{\hat 1}_\eta &\equiv & {\hat 1}_\eta(F),\ {\hat 0}_\rho\equiv {\hat 0}_\rho(F),\nonumber\\*
%1_\eta &\equiv & 1_\eta({\cal V}_M^F),\ 0_\rho\equiv 0_\rho({\cal V}_M^F),\label{eq10}
%\end{eqnarray}
%where $1_\eta$ and $0_\rho$ depend on ${\cal V}_M^F$ rather 
%than the full formal context $F$ since they are not affected by relabelling the objects.
This is the logic foundation provided by 
inspecting $\mu$ for its object partial ordering from
the object-attribute relationship 
embedded in the {\bf GCL} structure.
% can be utilised to exhibit how  
%the object-sets and generalised attributes being related by the formal context.
However, the conventional interest of
{\it rules of implication} are attribute based, where
the object reference is implicit.
%which develops logic relations of pure attribute-dependencies.
%Such a concern certainly hides the explicit object-labelling 
%in a consistent background.
Thus, one's primary concern 
here is the so-called {\it implication informative} above the {\bf GCL} framework, which was 
discussed originally in the traditional FCL framework \cite{GD86}. 

Note that for FCL an 
implication relation $A\rightarrow B$ 
is considered between the attribute sets
$A\subseteq M$ and $B\subseteq M$. 
Since the FCL in effect deals with the {\it conjunctions} of simple attributes in $M$, see Lemma~3.1~in~Ref.~\cite{LLJD12-1}, 
every FCL rule receives a corresponding rule in the {\bf GCL}:
\begin{equation}
A\stackrel{fcl}{\rightarrow} B\ \mbox{with}\ \lsmatr{A\in 2^M\\
B\in 2^M}\quad\mbox{corresponds to}\quad
\mu_1\rightarrow\mu_2\ \mbox{with}\ \lsmatr{\mu_1:=\prod A\ \in M^\ast\\
\mu_2:=\prod B\ \in M^\ast}.\nonumber
%\label{eq:implica_fcl_Gcl}
\end{equation}
Therefore, one is interested in the implication $\mu_1\rightarrow\mu_2$
which relates between the generalised attributes 
$\mu_1\in M^\ast$ and $\mu_2 \in M^\ast$
in the {\bf GCL} theory. In particular, according to Ref. \cite{GD86}, 
{$A\stackrel{fcl}{\rightarrow} B$} is a tautology 
(i.e. {\it not informative}) if $A\supseteq B$, which means $\mu_1=\prod A\leq \prod B=\mu_2$. Hence, if $\mu_1\not\leq \mu_2$ then the {\bf GCL} rule {$\mu_1\rightarrow\mu_2$} is informative.
%if $\mu_1\not\leq \mu_2$" 
%because $A\supseteq B$ iff $\prod A\leq \prod B$.\\
%However, let one employ the following refinement.\\
%Let, henceforth, the considerations be all based on 
%the {\bf GCL} subject to a given formal context $F(G,M)$.\\
%{\bf Definition~4.5}\\
%for {\bf GCL}
\dfn{
Consider the implication statement 
{$\mu_1\rightarrow\mu_2$} ($\mu_1$ implies $\mu_2$), where 
$\mu_1\in M^\ast$ and $\mu_2\in M^\ast$. 
\begin{itemize}
\item
If $\mu_1>\mu_2$,
{$\mu_1\rightarrow\mu_2$} is referred to as a 
{\it rule of {purely informative} implication} ({\bf RPII}).
\item
If $\mu_1\not\leq \mu_2$, {$\mu_1\rightarrow\mu_2$} is referred to as a 
{\it rule of {informative} implication} 
({\bf RII}).
\item
If {$\mu_1\rightarrow\mu_2$} is manifestly true ($\mu_1\leq\mu_2$) then it is referred to as a {tautology} ({\bf TT}), in the sense that the implication tells nothing new, and is re-denoted as {$\mu_1\implies \mu_2$}. 
\end{itemize}
Note that {\bf RPII} 
is a particular class of {\bf RII}'s in which no {\bf TT} 
is involved.} %%$\blacksquare$\\
%Here, an alternative interpretation for $\mu_1\rightarrow\mu_2$ is 
%that the object class possessing $\mu_1$ must be contained in the object class equipped with $\mu_2$.%\\ 
%{\bf Lemma 4.6}\\
\lem{
%Subject to the formal context $F(G,M)$, 
For the {\bf GCL}
subject to the formal context $F(G,M)$,
the rules of implication between two attributes in $M^\ast$
are well defined.
Explicitly, $\forall\mu_1\forall \mu_2\in M^\ast$, $\mu_1\rightarrow\mu_2$ can be identified as $\mu_1^R\subseteq \mu_2^R$: 
$\mu_1\leftrightarrow\mu_2$ ($\mu_1^R=\mu_2^R$)
is referred to as
the {\bf T1} rule;
$\mu_1\rightarrow\mu_2$ ($\mu_1^R\subset\mu_2^R$)
is referred to as
the {\bf T2} rule.
\begin{itemize}
\item
One may deduce {\bf T2} from {\bf T1} by {\bf TT}.
\item
Knowing all the {\bf RPII}'s suffices 
the full characterisation of {\bf RII}'s ({\bf Definition~4.5}).
\end{itemize}
}
%{\bf Proof}: \\
\prf{
Similar to the idea of $A\rightarrow A^{II}$ in Ref.~\cite{GD86},
the rule of implication $\mu_1\rightarrow\mu_2$ means that 
{the object class possessing $\mu_1$ 
must also be equipped with $\mu_2$}.
In other words, the object class possessing $\mu_1$ 
is included in the object class possessing $\mu_2$,
hence, $\mu_1^R\subseteq \mu_2^R$. 
Note that the {\bf T1} rule is a {\it bi-implication}.
With $\mu_1^R=\mu_2^R:=X$, 
both $\mu_1$ and $\mu_2$ are in $[X]_F$,
they belong to the common property
of the same object class and are thus regarded as equivalent.
For {\bf T2}, let $X_1=\mu_1^R$ and $X_2=\mu_2^R$,
where $X_1\subset X_2$.
Thus, $(\mu_1\cdot\mu_2)^R=X_1\cap X_2=X_1$, and $\mu_1\cdot\mu_2\in [X_1]_F$. Accordingly,
$gR\mu_1\implies gR(\mu_1\cdot\mu_2)$ entails that
the object possessing $\mu_1$ must also possess $\mu_2$. 
\begin{itemize}
\item
This is to show that $\mu_1^R\subset \mu_2^R$ implies $\mu_1\rightarrow\mu_2$ based on {\bf T1}.\\
%Deducing {\bf T2} from {\bf T1} is possible.
Since $\mu^R\in E_F\ \forall \mu\in M^\ast$,
if $\mu_1^R\subset \mu_2^R$ then  $\eta (\mu_1^R)<\eta(\mu_2^R)$
%$\left\lbrace \begin{smallmatrix}
%\eta (\mu_1^R)<\eta(\mu_2^R) \\
%\rho (\mu_1^R)<\rho(\mu_2^R)
%\end{smallmatrix}\right.$ 
by {\bf Proposition~2.1} (also cf. Lemma~3.14 of Ref.~\cite{LLJD12-1}), thus, $\eta(\mu_1^R)\implies \eta(\mu_2^R)$.
%$\left\lbrace \begin{smallmatrix}
%\eta(\mu_1^R)\rightarrow \eta(\mu_2^R) \\
%\rho(\mu_1^R)\rightarrow \rho(\mu_2^R)
%\end{smallmatrix}\right.$. 
In addition, with $\mu_1^R\subset \mu_2^R$
two {\bf T1} rules state that 
$\left\lbrace \begin{smallmatrix}
\mu_1\leftrightarrow\eta(\mu_1^R)\\
\mu_2\leftrightarrow\eta(\mu_2^R)
\end{smallmatrix}\right.$.
Therefore, based on {\bf T1}, $\mu_1^R\subset \mu_2^R$ implies
 $\mu_1 \rightarrow \mu_2$ ({\bf T2}). 
%Conversely, assume "$\mu_1\rightarrow\mu_2$" is {\bf T2}-rule, i.e. $\mu_1^R\subset \mu_2^R$.
%Then, "$\mu_1\rightarrow\mu_2$"
%further leads to 
%"$\left\lbrace \begin{smallmatrix}
%\mu_1\rightarrow\mu_1\cdot\mu_2\\
%\mu_1+\mu_2\rightarrow\mu_2
%\end{smallmatrix}\right.$", which are of the type {\bf T1}, since 
%$\left\lbrace \begin{smallmatrix}
%\mu_1^R=(\mu_1\cdot\mu_2)^R=\mu_1^R\cap\mu_2^R\\
%(\mu_1+\mu_2)^R=\mu_1^R\cup\mu_2^R=\mu_2^R
%1\end{smallmatrix}\right.$. %$\blacksquare$ 
\item
%Since {\bf T2} can be deduced from {\bf T1}, all the rules of implication can
%be determined from {\bf T2}.
Any rule {$\mu_1\rightarrow\mu_2$} 
where $\mu_2$ and $\mu_1$ are not ordered
is a {\bf RII}.
However, $\mu_1\rightarrow\mu_2$
is logically equivalent to $\mu_1\rightarrow \mu_1\cdot\mu_2$ 
because %with $\mu_1\rightarrow \mu_1\cdot\mu_2$ 
one may invoke the tautology $\mu_1\cdot\mu_2\implies \mu_2$ 
so as to recover $\mu_1\rightarrow \mu_2$. 
Apparently, $\mu_1\rightarrow \mu_1\cdot\mu_2$ is an {\bf RPII} since $\mu_1\cdot\mu_2<\mu_1$.
Therefore, knowing all the {\bf RPII}'s suffices 
the full characterisation of {\bf RII}'s.
%and is a {\bf T1} rule because $(\mu_1\cdot\mu_2)^R=\mu_1^R=\mu_2^R$ 
\end{itemize}}
%The above in effect asserts that the validity of "$\mu_1\rightarrow\mu_2$" 
%$\left\lbrace \begin{smallmatrix}
%\mu_1\in [X_1]\\
%\mu_2\in [X_2]_F
%\end{smallmatrix}\right.$ is based on $X_1\subseteq X_2$.
While {\bf T1} works in two directions, it is not always the case that both implications are informative.  
For instance, if {$\mu_1\rightarrow\mu_2$} is an {\bf RPII} 
then $\mu_1\leftarrow\mu_2$ is a {\bf TT} since $\mu_1>\mu_2$.
%Certainly, it can also happen that both directions are not {\bf RPII}s.
Likewise, $\mu_1\rightarrow\mu_2$ of the type {\bf T2} needs not be 
a {\bf TT} because it is not necessarily 
$\mu_1< \mu_2$ although $\mu_1^R\subset\mu_2^R$.
Moreover, none of {\bf T2} rules can be {\bf RPII} 
since  
{\it tautology} 
has been involved in deducing them from the {\bf T1} rules.
%Remarkably, $\mu_2$ needs not be unique subject to $\mu_1$ for both {\bf T1} and {\bf T2}. 
%Also, {\it not all} these $\mu_2$s end up with {\bf RII}.
%However, the rules of implication remain manageable.
%{\bf Proposition 4.7}\\
\prp{
All the rules of implication can be determined
in the following sense.
%"$\mu\rightarrow \eta(\mu^R)$" and "$\rho(\mu^R)\rightarrow \mu$" within Eq. (\ref{eq11})
%can be employed to determine all the rules of implication 
%essentially holding between any two attributes in $M^\ast$ 
\begin{itemize}
\item
%Any {\bf RII} $\mu_1\rightarrow \mu_2$ is expected 
%between some $\mu_1\in [X_1]_F$ and $\mu_2\in [X_2]_F$ 
%only when $X_1$ and $X_2$ is ordered.
%Thus, an {\bf RII} is either of {\bf T1} and {\bf T2} given in {\bf Lemma 4.6} 
%and all the rules of implication can be determined based on the {\bf RPII}s.
$\forall \mu \in M^\ast$, any {\bf RII} with respect to $\mu$ 
can be deduced from {$\mu\rightarrow \mu\cdot 1_\eta$}
by {\bf TT}. 
\item
{$\forall \mu \in M^\ast\ \mu\rightarrow \mu\cdot 1_\eta$} 
is equivalent to {$\forall \mu \in M^\ast\ \mu+ 0_\rho\rightarrow\mu$}
when implementing the rules of implication.
\end{itemize}
}
\prf{
Since knowing {\bf RPII} suffices 
the full characterization of {\bf RII}
and all the rules of implication can be deduced from the type {\bf T1} ({\bf Lemma~4.6}),
one is only interested in
the {\bf T1}/{\bf RPII}
$\forall \mu \in M^\ast$ written as\\
$\mu\rightarrow\nu_1\ \mbox{for}\ \nu_1 \in [\eta(\mu^R),\mu)$ and 
$\nu_2\rightarrow\mu\ \mbox{for}\ \nu_2 \in (\mu,\rho(\mu^R)]$.
\begin{itemize}
\item
%for any given attribute $\mu$ in $ M^\ast$, which are 
By {\bf Proposition 4.1},
 {$\mu\rightarrow \mu\cdot 1_\eta$} is equivalently to {$\mu\rightarrow \eta(\mu^R)$}.
%It suffices to show that "$\mu\rightarrow \mu\cdot 1_\eta$" can implement both the above expressions.
Since $\eta(\mu^R)$ is the lower bound of the interval $[\eta(\mu^R),\mu)$, 
the consequence of the disjunction-introduction {$\lrp{\mu\rightarrow \eta(\mu^R)}\implies \lrp{\forall \tau\in M^\ast\ (\mu\rightarrow \eta(\mu^R)+\tau)}$}
can exhaust all the possibilities of $\mu\rightarrow \nu_1$ for $\nu_1\in [\eta(\mu^R),\mu)$.
On the other hand, 
$\nu_2\cdot 1_\eta=\eta(\mu^R)$ for $\nu_2 \in (\mu,\rho(\mu^R)]\subseteq [\mu^R]_F$, rendering 
%{$\nu_2\rightarrow \nu_2\cdot 1_\eta$}
%become 
{$\nu_2\rightarrow \eta(\mu^R)$}.
Therefore, 
$\nu_2\rightarrow \mu$ for $\nu_2 \in (\mu,\rho(\mu^R)]$
since $\eta(\mu^R)\implies \mu$.  %$\blacksquare$
% {\it GE} that possessing the attribute $\mu$. ({\bf Proposition 4.1}) 
%However, $[\mu^R]_F=[\eta(\mu^R), \rho(\mu^R)]$, namely $\eta(\mu^R)\leq\mu\leq\rho(\mu^R)$ ({\bf Corollary 4.3}).
%Therefore, the smallest property (strictest condition)
%fulfilled by $\mu^R$ is $\eta(\mu^R)=\mu\cdot 1_\eta$, hence $\mu\rightarrow \mu\cdot 1_\eta$.
%Clearly, this becomes 
%tautology when $\mu$ happens to be $\eta(\mu^R)$ 
%In addition, the largest property (loosest condition) which implies $\mu$ is $\rho(\mu^R)=0_\rho +\mu$, 
%thus $\mu+0_\rho \rightarrow \mu$. Likewise, tautology occurs at $\mu=\rho(\mu^R)$.
\item
%The {\bf RPII} "$\mu +0_\rho\rightarrow \mu$" prescribes the objects in $\mu^R\in E_F$:
%An object in $\mu^R$ must also possesses $\mu +0_\rho$.
%However, $\mu +0_\rho\rightarrow \mu$ iff $\neg \mu \rightarrow\neg(\mu +0_\rho)$ which
%concerns the complementary set $(\neg \mu)^R\equiv\mu^{Rc}\in E_F$.
$(\forall \mu \in M^\ast\ \mu +0_\rho\rightarrow \mu) \iff  
(\forall \neg \mu \in M^\ast\ \neg \mu \rightarrow\neg(\mu +0_\rho))
\iff (\forall \neg \mu \in M^\ast\ \neg \mu\rightarrow\neg \mu \cdot 1_\eta)$\\
$\iff (\forall \mu\in M^\ast\ \mu \rightarrow \mu\cdot 1_\eta)$. %$\blacksquare$
%Hence, "$\forall \mu \in M^\ast,\ \mu +0_\rho\rightarrow \mu$" is equivalently "$\forall \neg \mu \in M^\ast,\ \neg \mu \rightarrow\neg(\mu +0_\rho)$".
%Moreover, $\neg (\mu +0_\rho)\equiv \neg \mu \cdot 1_\eta$.
%If $\mu^{Rc}$ does {\it not} possess $\mu+0_\rho$ then it does {\it not} possess $\mu$. 
%Subsequently, "$\forall \neg \mu \in M^\ast,\ \neg \mu\rightarrow\neg \mu \cdot 1_\eta $" is exactly the same as "$\forall \mu\in M^\ast,\ \mu \rightarrow \mu\cdot 1_\eta$". %$\blacksquare$
\end{itemize}
}
%It is obvious that the rule {$\mu\rightarrow \mu\cdot 1_\eta$} 
%{\it maximises} the difference 
%between $\mu$ and $\nu$ within all the choices of 
%$\nu\in [\mu\cdot 1_\eta,\mu)$ for the {\bf RPII} \q{$\mu\rightarrow \nu$}.
%The rule \q{$\mu\rightarrow \mu\cdot 1_\eta\equiv \eta(\mu^R)$}
%thus can be employed to deduce any
%\q{$\mu\rightarrow \nu$} with $\nu\in (\mu\cdot 1_\eta,\mu)$.
%Likewise, $\forall X\in E_F$, from $\rho(X)\rightarrow \eta(X)$
%one may obtain $\forall \mu \in [X]_F\ \mu\rightarrow \eta(X)$.
%
%It turns out that 
%the logic consequence resulted from the {\bf GCL} 
%can be completely implemented in 
%the $2^{n_F}$ {\it maximised} {\bf T1}/{\bf RPII}'s \q{$\rho(X)\rightarrow \eta(X)\ \forall X\in E_F$}, 
%up to the {\bf TT} based on the mathematical evidence
%{$X^1\subset X^2\iff \left\lbrace \begin{smallmatrix}
% \eta (X^1)<\eta(X^2) \\
% \rho (X^2)<\rho(X^2)
% \end{smallmatrix}\right.$} (cf. {\bf Lemma~4.6}). 
%Similar to the construction of {\it IIM} in Ref.~\cite{GD86},
%Clearly,
%the implementation 
%is that $\forall \mu_1\forall \mu_2\in [\eta(X),\rho(X)]$
%\q{$\mu_1>\mu_2$} gives a {\bf T1}/{\bf RPII} \q{$\mu_1\rightarrow\mu_2$}.
Therefore, by applying the formula {$\mu\rightarrow \mu\cdot 1_\eta$}
one is able to deduce all the logic implications 
implemented by the {\bf GCL}. 
%generating a {\bf RPII} from any given attribute, 
Moreover, 
two additional points remain crucial in concluding 
that the result at hand 
%the logic deduction based {\bf GCL} 
in fact avoids 
the tractability problem for implementing the logic content 
\cite{GW99,GD86,Kso04,KO08,Sb09,Df10,DS11,BK13} in FCL.
Firstly, the {\bf GCL}-based deduction is sufficiently general to include implication rules deduced from {FCL} and {RSL}.
Secondly, 
the unique formula {$\mu\rightarrow \mu\cdot 1_\eta$} 
%({\bf Proposition~4.7}) 
can furnish 
a systematic decision about whether any given rule
is supported by the formal context. 
\begin{itemize}
\item
From the perspective of Ref.~\cite{GD86}
$A\stackrel{fcl}{\rightarrow} B$ is furnished by $A^{I}\subseteq B^{I}$, while
one may have its analogue in RSL by  
identifying $A\stackrel{rsl}{\rightarrow} B$ with
$A^{\Diamond}\subseteq B^{\Diamond}$. 
Moreover, there are two equivalence relations inspired by Lemma~3.1~of~Ref.~\cite{LLJD12-1}: 
$A\stackrel{fcl}{\rightarrow} B$ is equivalent to $\prod A{\rightarrow} \prod B$;
$A\stackrel{rsl}{\rightarrow} B$ is equivalent to $\sum A{\rightarrow} \sum B$.
Both are particular cases of 
$(\mu_1\rightarrow \mu_2)\iff (\mu_1^R\subseteq\mu_2^R)$ ({\bf Lemma~4.6}) since
\begin{eqnarray}
A^I\subseteq B^I \iff \mu_1^R\subseteq\mu_2^R\ \mbox{with}\ \lsmatr{\mu_1=\prod A\\ \mu_2=\prod B}
&\because& \lsmatr{A^I=\bigcap_{m\in A} m^R=(\prod A)^R=\mu_1^R\\
B^I=\bigcap_{m\in B} m^R=(\prod B)^R=\mu_2^R},\nonumber\\
A^{\Diamond}\subseteq B^{\Diamond} \iff \mu_1^R\subseteq\mu_2^R\ \mbox{with}\ \lsmatr{\mu_1=\sum A\\ \mu_2=\sum B} 
&\because& \lsmatr{A^{\Diamond}=\bigcup_{m\in A} m^R=(\sum A)^R=\mu_1^R\\
B^{\Diamond}=\bigcup_{m\in B} m^R=(\sum B)^R=\mu_2^R},\label{eq:fclrsl_impl}
\end{eqnarray}
where $(A^{\Diamond},B^{\Diamond})$ and $(A^I,B^I)$ are referred to the variant forms given in Eq. (9) of Ref.~\cite{LLJD12-1}.
%is the analogue of 
%$A\stackrel{fcl}{\rightarrow} B$ with
%$A^{I}\subseteq B^{I}$.
For instance, let one inspect the content of 
{$\rho(\lbrace 1,2\rbrace) > c
> \eta(\lbrace 1,2\rbrace)$} 
in Fig. \ref{fig:rslfcl_gcl}
by employing the fact
$c^R=\lbrace 1,2\rbrace$ is a common extent of {FCL} and {RSL},
where implications of both types can be easily exhibited.\\
%Consider thus 
%$\rho(\lbrace 1,2\rbrace)={c+d+(a+\neg b+\neg e)(\neg a+b+e)(\neg a+\neg b+\neg e)}$ and $\eta(\lbrace 1,2\rbrace)={a\neg bc(de+\neg d\neg e)}$.
%\begin{equation}
%\rho(\lbrace 1,2\rbrace)=\underbrace{c+d+(a+\neg b+\neg e)(\neg a+b+e)(\neg a+\neg b+\neg e)}_{\rho(\lbrace 1,2\rbrace)}\rightarrow \underbrace{a\neg bc(de+\neg d\neg e)}_{\eta(\lbrace 1,2\rbrace)}=\eta(\lbrace 1,2\rbrace).%\label{eq12}
%\end{equation}
For $c \rightarrow
\eta(\lbrace 1,2\rbrace)$,  
the rule $c\rightarrow a\neg bc(de+\neg d\neg e)$ is 
of the type {\bf T1}/{\bf RPII} but its equivalent statement 
$c\rightarrow a\neg b(de+\neg d\neg e)$ is simply a {\bf T2}-rule, 
which is due to
$c^R=\lbrace 1,2\rbrace\subset \lbrace 1,2,5\rbrace=(a\neg b(de+\neg d\neg e))^R$.
On the other hand,
{$\lrbr{c}\stackrel{fcl}{\rightarrow} \lrbr{a}$} 
is due to $\lbrace c\rbrace^I=\lrbr{1,2}\subset \lrbr{1,2,5,6}=\lbrace a\rbrace^I$ \cite{GD86},
in addition, 
%{$\lrbr{c}\stackrel{fcl}{\rightarrow} \lrbr{a}$} is equivalent to 
{$\lbrace c\rbrace\stackrel{fcl}{\rightarrow}\lbrace a,c\rbrace$}.
Based on Eq. (\ref{eq:fclrsl_impl}), 
$\lrbr{c}\stackrel{fcl}{\rightarrow} \lrbr{a}$ 
becomes the {\bf T2}-rule
{$c\rightarrow a$} and
{$\lbrace c\rbrace\stackrel{fcl}{\rightarrow}\lbrace a,c\rbrace$}
becomes the {\bf T1}-rule
{$c\rightarrow ac$}, 
both are the consequences of $(c\rightarrow a\neg bc(de+\neg d\neg e))\implies (c\rightarrow ac)$.\\
For $\rho(\lbrace 1,2\rbrace) \rightarrow c$,
the rule {$c+d+(a+\neg b+\neg e)(\neg a+b+e)(\neg a+\neg b+\neg e)\rightarrow c$} 
%does not result in any implication in the frame of FCL. However, 
%$(c+d+(a+\neg b+\neg e)(\neg a+b+e)(\neg a+\neg b+\neg e)\rightarrow c)$
can imply $(d\rightarrow c)$ and $(c+d\rightarrow c)$.  
Discarding the disjunction symbol, it turns out that 
$\lbrace d\rbrace\stackrel{rsl}{\rightarrow} \lbrace c\rbrace$ and
$\lbrace c,d\rbrace\stackrel{rsl}{\rightarrow} \lbrace c\rbrace$, 
which are % is then understood as $\lbrace c\rbrace^{\Diamond\Box}\rightarrow\lbrace c\rbrace$ 
what one could have achieved in RSL 
\footnote{To our knowledge, the 
discussion about such relations has not yet been found in the literature.}:
$\lbrace d\rbrace\stackrel{rsl}{\rightarrow}\lbrace c\rbrace$
is based on $\lbrace d\rbrace^\Diamond\subseteq\lbrace c\rbrace^\Diamond$; 
$\lbrace c,d\rbrace\stackrel{rsl}{\rightarrow}\lbrace c\rbrace$
is based on $\lbrace c,d\rbrace^\Diamond\subseteq\lbrace c
\rbrace^\Diamond$. 
In principle, 
identifying $A\stackrel{rsl}{\rightarrow} B$ with
$A^{\Diamond}\subseteq B^{\Diamond}$ 
is the reasonable analogue of 
$A\stackrel{fcl}{\rightarrow} B$ with
$A^{I}\subseteq B^{I}$.
%To the reason why 
%the implication rules 
%obtained from {FCL} and {RSL}
%can be implemented in the {\bf GCL},
%it is {\bf Proposition~2.2} that states
%\[
%\mbox{{\bf GCL}}\supset
%{\left\lbrace \begin{array}{ccccc}
%\mbox{{gFCL}}&\supset&\mbox{FCL}&\vdash& \mbox{the implication rules in Ref. \cite{GD86}}\\
%\mbox{{gRSL}}&\supset&\mbox{RSL}&\vdash& \mbox{the implication rules that could have been developed for RSL\ 
%}\\
%\mbox{{cgRSL}}&\supset&\mbox{{\it po-rsl}}&:&\ldots\\
%\ldots&&&\vdash&\ldots
%\end{array}\right.
%}, 
%
%\]
%where $\vdash$ is to note 
%Basically, the nodes pertaining to FCL and RSL are found to be recoverable in the {\bf GCL}
%(Proposition 3.15 \cite{LLJD12-1}, Fig. \ref{fig:rslfcl_gcl}), while
%"$\ldots$" are referred to as certain 
%{\it possibly-existing yet-undiscovered} structures which are less obvious.
%However, 
%one wont be interested in what
%each of these concept lattices 
%would separately supply to one's knowledge about logics
%because
%{\it the most general case happens to be the one which is most easy to be implemented}:
%For {\bf GCL}, listing out all the {\bf RPII}s is simply a tedious task.\\
\item
Note that it is not practical
to list out all the possible {\bf RII}'s.
Instead, the {\bf GCL} provides criteria to determine 
whether a logic implication %assumed within any attribute pair
is allowable by the formal context.
Since 
$\mu_1\rightarrow\mu_2$
is defined by $\mu_1^R\subseteq \mu_2^R$,
which entails $\eta(\mu_1^R)\leq\eta(\mu_2^R)$ (Proposition~3.14~in~Ref.~\cite{LLJD12-1}),
the {$\mu_1\rightarrow\mu_2$}
is an implication allowable
by the formal context 
iff $\mu_1\cdot 1_\eta\implies\mu_2\cdot 1_\eta$. 
%In effect, $F(G,M)$ then proposes 
%through the theory that 
Since $1_\eta$ is logically true subject to the formal context,
it always coexists with any other attribute.
In other words, one requires
$\mu_1\rightarrow\mu_2$ to be true under the condition $1_\eta$:
\begin{equation}
\lrp{1_\eta\rightarrow \lrp{\mu_1\rightarrow\mu_2}}\iff 
\lrp{\mu_1\cdot 1_\eta\rightarrow\mu_2\cdot 1_\eta}.
\label{eq:c_imply_or_not}
\end{equation}
Moreover, if $\mu_1\rightarrow\mu_2$ is allowable by $1_\eta$ then one has ${\mu_1\cdot 1_\eta\implies\mu_2\cdot 1_\eta},\ \mbox{i.e., }\mu_1\cdot 1_\eta\leq\mu_2\cdot 1_\eta$.
For example, by employing 
Eq.~(\ref{eq:etadnf}) for
the $1_\eta$ obtained in Fig. \ref{fig:rslfcl_gcl}, 
one has
\begin{eqnarray}
c\rightarrow a\quad &\because &\quad \underbrace{a\neg bc(de+\neg d\neg e)}_{c\cdot 1_\eta}<\underbrace{a\neg bc(de+\neg d\neg e)+a\neg c\neg d(be+\neg b\neg e)}_{a\cdot 1_\eta}\nonumber\\
\neg e\rightarrow a+\neg b\neg d &\because& \quad\neg e\cdot 1_\eta=
a\neg bc\neg d\neg e+a\neg b\neg c\neg d\neg e<(a+\neg b\neg d)\cdot 1_\eta\nonumber\\
d\leftrightarrow ace\quad &\because & \quad\quad d\cdot 1_\eta=a\neg bcde=ace\cdot 1_\eta\nonumber\\
b+cd\leftrightarrow e &\because & \quad (b+cd)\cdot 1_\eta=a\neg bcde+\neg ab\neg c\neg de+ab\neg c\neg de=e\cdot 1_\eta\nonumber\\
c\left\lbrace \begin{smallmatrix}
\not \rightarrow \\
\not \leftarrow
\end{smallmatrix}\right.e\quad &\because &c\cdot 1_\eta=a\neg bc(de+\neg d\neg e)
\left\lbrace \begin{smallmatrix}
\not > \\
\neq\\
\not <\end{smallmatrix}\right.
a\neg bcde+\neg ab\neg c\neg de+ab\neg c\neg de=e\cdot 1_\eta
\label{eq:imply_or_not}
\end{eqnarray}\end{itemize}

\begin{figure}  
\includegraphics[scale=0.6,angle=90]{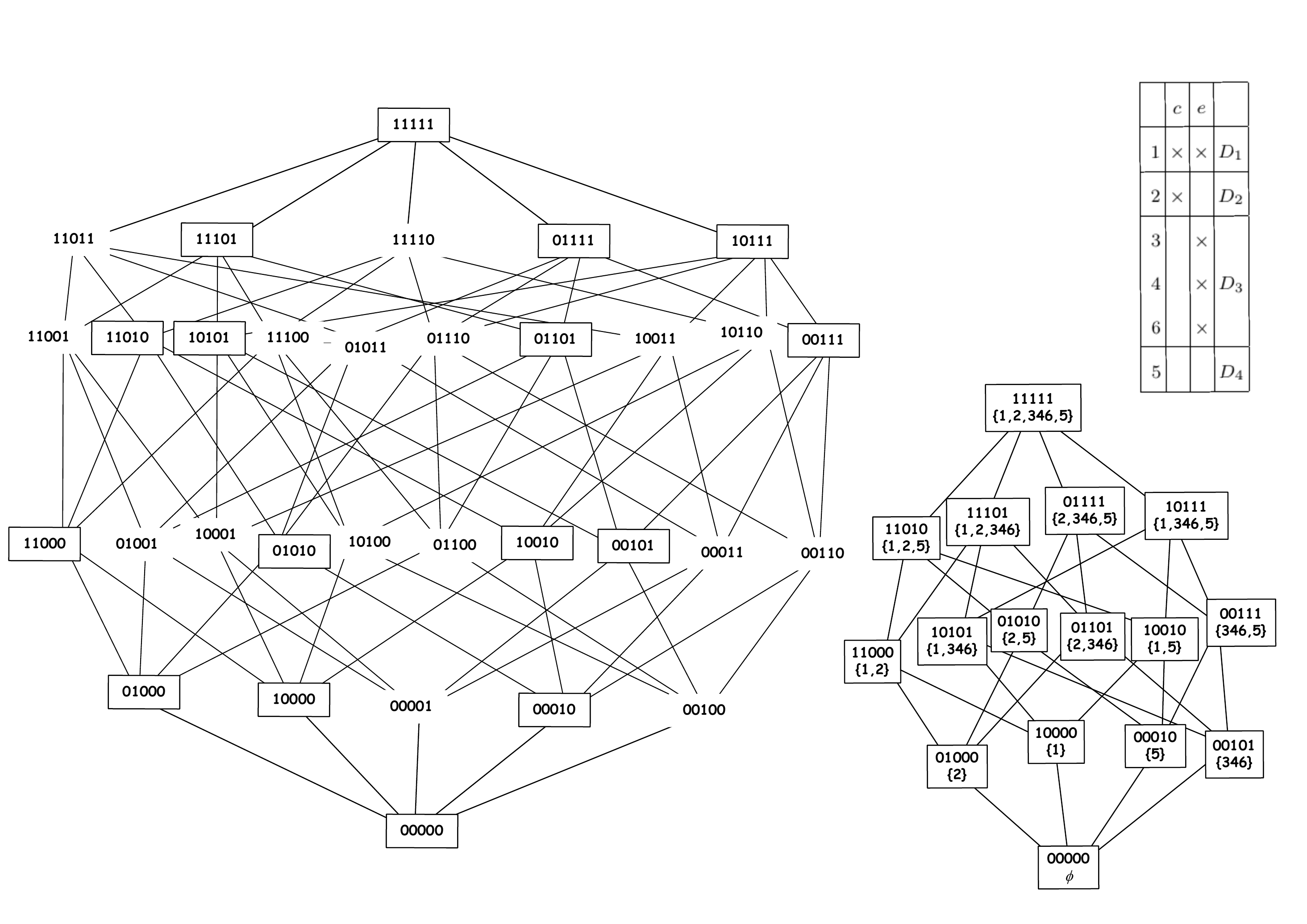}
\caption{
A {\bf GCL} obtained after removing $\lbrace a,b,d\rbrace$ 
from $M=\lbrace a,b,c,d,e\rbrace$ in the formal context $F(G,M)$ of 
Table \ref{table:formal_scheme}, say 
$F_{res}(\lbrace 1,2,3,4,5,6\rbrace,\lbrace c,e\rbrace)$\\
Various object classes collapse into one, 
because the remaining attributes are insufficient to preserve 
the categorizations offered by Fig. \ref{figure:context_venn}. 
The $16$ boxed general extents mark the residual classification on the original {\bf GCL} to form a degenerate one.  
Notably, $\rho(X)=\eta(X)\ \forall X\in E_{F_{res}}$, e.g., 
$\eta(\lbrace 2,5 \rbrace)=[0101]_\eta=c\neg e+\neg c\neg e=\neg e=(\neg c+\neg e)(c+\neg e)
=[0101]_\rho=\rho(\lbrace 2,5 \rbrace)$.} 
\label{fig8}
\end{figure}

%as "$1_\eta\rightarrow (\mu_1\rightarrow\mu_2)$" 
Another interesting point is about the limit
at which the contextual truth $1_\eta$ becomes 
the {\it real} logical truth ${\bf 1}$. 
Under consideration is thus the {\it degenerate} {\bf GCL} which emerges from a {\it degenerate} formal context as follows.
\prp{
%{\bf Proposition~4.8}\\
Subject to $F(G,M)$, 
the following statements are all equivalent, which defines the {\it degenerate} formal context:
{\bf S1.}
$n_F=\log_2^{|M^\ast|}$.
%i.e., the number of discernible (by $F(G,M)$) subsets coincides with 
%$rank(M)$ (cf. Eq. (\ref{eq:e_k})),
%which is the maximal number of object-classes one may distinguish based on $M$. 
{\bf S2.}
$\forall X\in E_F\ \rho(X)=\eta(X)$. 
%for all $X$ being allowable extent of $X$.
{\bf S3.}
$1_\eta={\bf 1}$ and $0_\rho={\bf 0}$.
{\bf S4.} $\forall\mu\in M^\ast\ [\mu^R]_F=\lrbr{\mu}$.}
%{\bf S5} Every $\mu\in M^\ast$ is a {\it Gfcp}.\\
%\end{description}
%{\bf Proof}:\\
\prf{
Consider {${\bf S1}\implies {\bf S2}\implies {\bf S3}\implies {\bf S4}\implies {\bf S1}$},
which entails {${\bf S1}\iff {\bf S2}\iff{\bf S3}\iff {\bf S4}$,
as follows.\\
{\bf S1}$\implies${\bf S2}}:
If $n_F=\log_2^{|M^\ast|}$ 
%the number of nodes of {\bf GCL} is equivalent to $|M^\ast|$ and 
then $|[X]_F|=1$ ({\bf Corollary~4.4}).  
%Corollary 4.4 $|[X]_F|=1$
%Since $\rho(X)\neq \rho (X^\prime)$ iff $X\neq X^\prime$ (Lemma 3.14 \cite{LLJD12-1}), 
%the $2^{\log_2^{|M^\ast|}}$ different {\it Grsp}s exhaust $M^\ast$.
Moreover, $\rho(X)\geq \eta(X)$ and $\lsmatr{\eta(X)\in [X]_F\\
\rho(X)\in [X]_F}$, therefore, $\rho(X)=\eta(X)$.\\ 
%because no extra 
%freedom, which can accomplish $\rho(X)\neq \eta(X)$ at any $X$ in $M^\ast$, is left .
{\bf S2}$\implies${\bf S3}:
$\rho(X)=\eta(X)$, thus,
$\forall \mu\in M^\ast\ \mu\cdot 1_\eta=\mu +0_\rho$ ({\bf Proposition 4.1}),
which renders $\mu \geq\mu\cdot 1_\eta=\mu +0_\rho\geq \mu$ by Eq.~(\ref{eq:mu_upper_lower}).
Therefore,
$\forall \mu\ \mu\cdot 1_\eta=\mu=\mu +0_\rho$ which implies $1_\eta={\bf 1}$ and $0_\rho={\bf 0}$.\\
{\bf S3}$\implies${\bf S4}:
$\mu+0_\rho=\mu+{\bf 0}=\mu\cdot {\bf 1}=\mu\cdot 1_\eta=\mu\ \therefore [\mu^R]_F=[\mu+0_\rho,\mu\cdot 1_\eta]=[\mu,\mu]=\lrbr{\mu}$.\\
%\item [S4$\Rightarrow$S5]
%If every $\mu\in M^\ast$ is a {\it Grsp} then every $\neg \mu \in M^\ast$ (equivalently, every $\mu^\prime\in M^\ast$) is a  {\it Gfcp}.
{\bf S4}$\implies${\bf S1}:
$\forall X_i\forall X_j \in E_F\ [X_i]_F\cap [X_j]_F=\emptyset$ iff $X_i\neq X_j$ ({\bf Proposition~2.1}).
Moreover, with {\bf S4}, 
$\forall\mu\in M^\ast$, $\mu^R$ can be identified as some 
$X\in E_F$, which is given by $\lrbr{\mu^R\mid\mu\in M^\ast}$.
Therefore, $\forall X\in E_F\ |[X]_F|=1=2^{\log_2^{|M^\ast|}-n_F}$, which implies $\log_2^{|M^\ast|}=n_F$.
} 
 %$\blacksquare$\\
%{\bf Corollary 4.9}\\
\crl{
Subject to a degenerate formal context, say $F_{\cal D}(G,M)$, 
$\forall \mu\forall \mu^\prime
\in M^\ast$ $\mu\leq \mu^{\prime}$ iff $\mu^R{\subseteq} (\mu^{\prime})^R$.
}
%{\bf Proof}: \\
\prf{
One now proceeds the proof in two parts.\\
Firstly,
$ \mu= \mu^{\prime}\stackrel{F_{\cal  D}}{\iff}\mu^R= (\mu^{\prime})^R$:
It is clear that $\mu=\mu^{\prime}\implies\mu^R= (\mu^{\prime})^R$.
Reversely, if $\mu^R= (\mu^{\prime})^R$ then 
$\lrbr{\mu}=[\mu^R]_{F_{\cal  D}}=[(\mu^{\prime})^R]_{F_{\cal  D}}=\lrbr{\mu^{\prime}}$, i.e. $\mu= \mu^{\prime}$.
Therefore, $\mu= \mu^{\prime}\iff\mu^R{=} (\mu^{\prime})^R$.\\ Secondly,
$\mu< \mu^{\prime}\stackrel{F_{\cal  D}}{\iff}\mu^R\subset (\mu^{\prime})^R$:
With $\mu= \mu^{\prime}\iff\mu^R{=} (\mu^{\prime})^R$, the formula {$\mu< \mu^{\prime}\implies \mu^R{\subseteq} (\mu^{\prime})^R$} (Lemma~2.10~in~Ref.~\cite{LLJD12-1}) is reduced to 
%However, $\mu\neq \mu^{\prime}\Rightarrow \mu^R\stackrel{F_{\cal  D}}{\neq} (\mu^{\prime})^R$, 
{$\mu< \mu^{\prime}\implies \mu^R{\subset} (\mu^{\prime})^R$}.
Reversely, assume $\mu^R{\subset} (\mu^{\prime})^R$ but $\mu\not < \mu^{\prime}$. Then, 
$\mu^R\cap (\mu^{\prime})^{Rc}=(\mu\cdot\neg \mu^{\prime})^R=\emptyset $ 
but $\mu\cdot\neg \mu^{\prime}\neq {\bf 0}$, which is contradictory since 
$\mu\cdot\neg \mu^{\prime}\in [\emptyset]_{F_{\cal D}}$ while $[\emptyset]_{F_{\cal D}}=\lbrace {\bf 0} \rbrace$. 
Therefore, $\mu^R{\subset} (\mu^{\prime})^R\iff\mu< \mu^{\prime}$.
} 
%$\blacksquare$\\
Note that the occurrence of degenerate {\bf GCL} is not as rare as 
one might have anticipated. For instance, by removing the attributes $a,b$ and $d$ from $F(G,M)$ 
in Table \ref{table:formal_scheme}, where $M$ 
reduces to $\lbrace c, e\rbrace$,
one will end up with $n_F=2^{|M|}=4$. 
As is depicted in Fig. \ref{fig8}, the resultant degenerate {\bf GCL} then comprises $2^4$ nodes embedded in the original one, 
in which $1_\eta=ce+c\neg e+\neg ce+\neg c\neg e\equiv{\bf 1}$ 
and $0_\rho=(\neg c+\neg e)(\neg c+e)(c+\neg e)(c+e)\equiv{\bf 0}$.
However, still more instructive 
is that to 
each formal context one may   
associate a degenerate formal context, 
which exhibits all the attribute freedom 
and can thus serve as the reference context for analysing 
the re-parametrisation for the general concept lattice.
%
%all the attributes in $M^\ast$ occur and are related 
%as if they were depicted in the conventional (free) Venn diagram of $M$. 
%the set-relations among $m^R$s  for $m\in M$ exactly follow the Venn diagram of $M$.
%Then, $[G]_F=\lbrace {\bf 1}\rbrace$ and $[\emptyset]_F=\lbrace {\bf 0}\rbrace$;
%$1_\eta=\eta (G)=\prod [G]_F={\bf 1}$ and $0_\rho=\rho (\emptyset)=\prod [\emptyset]_F={\bf 0}$.  
%Subsequently, $\mu\cdot 1_\eta=\mu\cdot {\bf 1}=\mu=\mu+{\bf 0}=\mu +0_\rho$ which means $\rho(X)=\eta(X)$ at any node of the %{\bf GCL} by {\bf Theorem 3.5}.
%$|M^\ast|$
%$a\neg bcde+a\neg bc\neg d\neg e+                      a\neg b\neg c\neg d\neg e                $
%$a\neg bcde+a\neg bc\neg d\neg e+\neg ab\neg c\neg de+a\neg b\neg c\neg d\neg e+ab\neg c\neg de$ 
%                                                    (a+\neg b+c+d+\neg e)                (\neg a+\neg b+c+d+\neg e)$
%(\neg a+b+\neg c+\neg d+\neg e)(\neg a+b+\neg c+d+e)(a+\neg b+c+d+\neg e)(\neg a+b+c+d+e)(\neg a+\neg b+c+d+\neg e)$. 

%$[01011]_\rho=(\neg a+b+\neg c+\neg d+{\bf \neg e})(a+\neg b+c+d+{\bf \neg e})$ 
%$[01011]_\eta={\bf a}\neg bc\neg d\neg e+{\bf a}\neg b\neg c\neg d\neg e+{\bf a}b\neg c\neg de$ 1111

\section{discussion}\label{five}
One has demonstrated in this paper the merits of {\bf GCL} 
with two insightful features, 
which are the generality and the tractability. 
For the generality feature,
it is shown that
the {\bf GCL} incorporates the conventional FCL and RSL
from both the perspectives of {\it lattice structure} and of 
{\it logic content}. 
From the {lattice-structure} perspective, the {\bf GCL} furnishes a comprehensive categorisation for whatever 
distinctive object classes (general extents) based on $F(G,M)$, where 
every attribute in $M^\ast$ essentially pertains to 
a definite general intent, cf. {\bf Proposition~2.1}.
The {\bf GCL} turns out to be the foundation of 
various generalised concept 
lattices ({\bf Proposition~2.2}) 
such as the generalised versions of 
the FCL and RSL.
In practice, all the nodes of the FCL and the RSL 
are identifiable on the {\bf GCL}, as can be explicitly worked out in 
Fig.~\ref{fig:construct_irreduicble} and  \ref{fig:rslfcl_gcl}.
%Note that the goal of approaches via concept lattice should not 
%be {\it to draw the lattices explicitly}.
%Typically, one may be asked of 
%finding a particular object class 
%by means of any of its associating attributes 
%subject to the 
%and
%e.g., on the {\bf GCL} any object 
%endowed with the attribute $\mu \in M^\ast$ must be classified into 
%the general extent $[\mu^R]_F$ ({\bf Proposition~4.1}).
%It is in such a manner that 
%
%It is more realistic than {\it to draw the lattice} is that 
%one can employ the resultant categorisation to respond on 
%a logic request such as finding a particular object class
%by means of part of its associating property ({\bf Proposition~4.1}).
%The {\bf GCL}
%furnishes logic deductions in such manner since
%any {\it implication} that relates 
%two different features is interpreted as a
%{\it set-inclusion} relation between 
%two object classes which respectively possess these two features.
%It is clear that the {\bf GCL}
%essentially comprises the FCL and RSL as 
%sub-parties.
%For {\bf GCL}, the rules of implication are in fact 
%more general than
%the generalisations 
%from the ones appearing in the FCL and the RSL.
From the {logic-content} perspective,
the logic implication extracted from the {\bf GCL}
is concerned with
the implication relations of the type
$\mu_1\rightarrow\mu_2$ ($\mu_1,\mu_2\in M^\ast$)
%defined in the perspective of {\bf GCL}
where the FCL- and RSL- based implications emerge 
as particular cases due to Eq.~(\ref{eq:fclrsl_impl}). 

%\[
%A\stackrel{fcl}{\rightarrow} B
%\iff 
%\mu_1\rightarrow\mu_2\ \mbox{with}\ \lsmatr{\mu_1=\prod A\\ \mu_2=\prod B},\quad
%A\stackrel{rsl}{\rightarrow} B \iff \mu_1\rightarrow\mu_2\ \mbox{with}\ \lsmatr{\mu_1=\sum A\\ \mu_2=\sum B}. 
%\]
%in contrast to the implication 
%applied to a pair of attribute-sets 
%$A,B\subseteq M$ (or $A,B\in 2^M$)
%By comparison, "$A\rightarrow B$" for FCL and RSL is respectively understood 
%as the subclass 
%$\mu_1:=\prod A\rightarrow \mu_2:=\prod B$
%and $\mu_1:=\sum A\rightarrow \mu_2:=\sum B$
%for {\bf GCL}.
%Clearly, 
%the FCL can be formulated 
%in terms of 
%the {conjunctions} of attribute-entities in $M$
%and the RSL is in terms of {disjunctions}
%. (Lemma~3.1~in~Ref.~\cite{LLJD12-1}).
%\begin{itemize}
%\item 
%\end{itemize}
For the tractability feature,
%{\bf For its tractability}:
%The construction of {\bf GCL}
%and the 
%logic deduction based on
%{\bf GCL} are shown to be extremely tractable.\\
both constructing the lattice structure
and implementing the logic content
are tractable.\\
To construct the lattice, the {\bf GCL} 
develops a Hasse diagram of $2^{n_F}$ nodes, where
each node is referred to as a general concept
comprising a distinct 2-tuple given in terms of
general extent and general intent.
The $2^{n_F}$ general extents  
appear to be all possible unions of the smallest subsets 
discernible by the formal context, 
see Proposition~3.5 in Ref.~\cite{LLJD12-1}, hence, no additional effort
is needed for selecting them out.
The general intents are 
$2^{n_F}$ {\it disjoint closed} sub-intervals of $M^\ast$ 
({\bf Proposition~2.1}, {\bf Corollary~4.4}) 
with the constant cardinality $2^{rank(M)-n_F}$. % (cf. Eq. (\ref{eq:e_k})).
The expression of a general concept is stated as 
%(general extent, general intent)
$(X, [X]_F)$, where 
$[X]_F=[\eta(X),\rho(X)]$, namely,  
$\eta(X)$ and $\rho(X)$ are respectively the lower and upper bounds
of $[X]_F$, see {\bf Proposition~2.1}.
The construction of {\bf GCL} is as tractable as 
listing out the formal context in an arbitray order since
it is fully characterised by means of 
the $\eta$-representation ${\Upsilon}_\eta$ 
(or $\rho$-representation ${\Upsilon}_\rho$, see {\bf Definition~3.4}), 
which can be completed by a single glance of the formal context, 
see {\bf Proposition~3.1}.
All the general concepts can then be read out on-demand
from ${\Upsilon}_\eta$ (${\Upsilon}_\rho$).
Note that based on {\bf Proposition~3.3}
any of the components in the triplet 
$(X,\rho(X),\eta(X))$ %of Eq.~(\ref{eq:g_concept_23})
will determine the other two, as can be illustrated 
by means of simple binary masks,
see Eq.~(\ref{eq:BX_k}).
% and Fig.~\ref{fig:Calculate_GfcpGrsp}.

In determining the logic content,
the {\bf GCL} suggests to 
implement the implication relations via
the entailment of the {\it lower bound} property.
Note that any object set carrying a definite property
should be categorised  into  a definite general extent 
because
every attribute in $M^\ast$ essentially belongs to 
a definite general intent ({\bf Proposition~2.1}).
Such an implementation is tractable since  
the single formula 
$\forall \mu\in M^\ast\ \mu\ \rightarrow \mu\cdot 1_\eta$ ({\bf Proposition~4.7})
suffices to present all 
the rules of informative implication
where $1_\eta$ is the contextual truth
obtained by summing all the components of the $\eta$-representation 
({\bf Definition~3.4}).
Conjugately, the formula can be restated as 
$\forall \mu\in M^\ast\ \mu+ 0_\rho\ \rightarrow \mu$
with the contextual falsity $0_\rho\equiv\neg 1_\eta$, 
which 
turns out to
implement 
the implication relations 
via
the entailment of 
%the {\it lower bound} property.
%turns out the entailment via 
the {\it upper bound}
property. 
%of the object class.
% which an object set with a given property
%should be grouped to.
Note that either of the formulas $\lsmatr{\mu\ \rightarrow \mu\cdot 1_\eta\\
\mu+ 0_\rho\ \rightarrow \mu}$
is capable of determining all the implication relations 
based on the formal context, including those
which could be deduced from the FCL and RSL
since both $A\stackrel{fcl}{\rightarrow} B$ 
and $A\stackrel{rsl}{\rightarrow} B$
can be interpreted as particular cases for 
$\mu_1\rightarrow \mu_2$ by Eq.~(\ref{eq:fclrsl_impl}).
On the other hand, 
$\lsmatr{A\stackrel{fcl}{\rightarrow} B\\
A\stackrel{rsl}{\rightarrow} B}$ can 
by no means generate all the possible implications
%one can generate 
from the formal context.
Obviously, an expression 
like \q{$\neg e\rightarrow a+\neg b\neg d$} in Eq.~(\ref{eq:imply_or_not}) 
can be identified neither
with $A\stackrel{fcl}{\rightarrow} B$ nor with 
$A\stackrel{rsl}{\rightarrow} B$.

The logic reasoning based on the {\bf GCL} 
is in fact rather intuitive.
All the attributes $[X]_F$ possessed by the same object class $X$ are regarded as equivalent.
For any $\mu\in M^\ast$, there is a {\it bi-conditional} equivalence 
{$\forall \nu\in [\mu^R]_F\ \mu\leftrightarrow \nu$} that
corresponds to the {\bf T1}-rules ({\bf Lemma~4.6})
from which one may determine all the rules by incorporating tautologies.
While the explicit object reference is ignored here,
logic relations only refer to the contextual Venn diagram ${\cal V}^F_M$,
%This coincides with the setting that 
$\mu_1\rightarrow \mu_2$ gets its interpretation via
$\mu_1^R\subseteq \mu_2^R$. %({\bf Lemma~4.6}).
In general, the set relation between 
$\mu_1^R$ and $\mu_2^R$ suggests
an ordering on ${\cal V}^F_M$
that further determines 
whether $\mu_1\rightarrow \mu_2$ is an allowable implication,
see Eq.~(\ref{eq:c_imply_or_not}) also cf. Proposition~3.14 in 
Ref.~\cite{LLJD12-1}. 
Note that any attribute
in effect serves as a logic statement that asserts 
property on a definite subject.
In particular, the attribute $1_\eta=1_\eta^F=1_\eta({\cal V}^F_M)$ 
exhibits a logic condition 
that governs the rules  
of implication %between any attribute can be established 
according to ${\cal V}^F_M$.  
%Now that $1_\eta$ is an attribute and stands for a logic condition, 
Hence, every single attribute enters as a logic statement 
that asserts property and then can be employed
as a logic condition 
%on assuming an underlying contextual Venn diagram 
%${\cal V}^{lc}_M$ 
by means of  $1_\eta({\cal V}^{lc}_M)$.
The idea to deal with the statements of pure attribute type
then brings about a {\it simplified} 
algebraically manipulable reasoning called 
the primary deduction system \cite{LLJD12-3} where the 
logical {\bf OR}, {\bf AND}, {\bf NOT} and implication 
among statements can be realised by the 
Boolean disjunction, conjunction and negation operators among attributes.
Indeed, the primary deduction system   
readily suffices to provide an efficient reasoning tool 
that leads to non-trivial applications, e.g. solving certain well-known puzzles. Moreover, the rules of classical logic 
are found to be  true in the primary deduction system 
since the Hilbert axioms in Ref.~\cite{Hd27}
all appear to be %logically
manifestly valid.
 %with sole attribute-typed statements.
It should however be remarked that 
the primary deduction system with pure attribute-type statements
could not be satisfactory 
and is coined to be na{\"i}ve
as it contradicts one's intuitions,
Ref.~\cite{LLJD12-4} thus strives to incorporate 
novel syntax in order to resolve such counterintuitive issue. 
Another point is that 
the primary deduction system is not expressive enough, as opposed 
to the conventional reasoning process, therefore,
one has to  look forward further to a comprehensive  
algebraically manipulable deduction \cite{LLJD12-5}. 

%To close this paper
In addition, the degenerate formal 
context ({\bf Proposition~4.8})  
describes a mathematical limit 
at which the number of object classes 
discernible by the formal context is exhausted.  
Notably, the degenerate formal 
context gives rise to a degenerate {\bf GCL} 
in which the condition $\rho(X)=\eta(X)$ 
reduces every general intent into 
one sole member of $M^\ast$ ({\bf Corollary~4.4}). 
The contextual Venn diagram for a degenerate formal context 
$F_{\cal D}$ in fact coincides with the conventional Venn diagram, 
say ${\cal V}^{F_{\cal D}}_M={\cal V}_M$, by means of
$1_\eta^{F_{\cal D}}={\bf 1}$.
Indeed, the implication formula  
({\bf Proposition~4.7}), when applying to the degenerate {\bf GCL}, 
concludes that $\forall \mu\in M^\ast\ \mu\rightarrow\mu$ 
which proposes no interesting implications 
and thus becomes less appealing as a practical categorisation.
Nevertheless, the degenerate formal context 
can serve as theoretical referential system.
To each formal context, a {\it referential context} 
can be designed to provide a basis convention,
%which the categorisation can work with,
by revealing the freedom 
of the generalised attribute system 
as was stated in Eq.~(\ref{eq:e_k}).
In practice, the {referential context} $F_{\cal D}(G\cup G_f,M)$ 
is a degenerate formal context
by appending $G_f$ to $F(G,M)$, a set of $rank(M)-n_F$
{\it fictitious} objects (hence, $n_{F_{\cal D}}=rank(M)$), 
as a means to expose the attribute freedom
corresponding to the properties
not carried by the existing objects.  
In Ref.~\cite{LLJD12-3}, 
it will be shown the referential 
context 
is instructive to illustrate the extensive structure of {\bf GCL}.
Moreover, $F_{\cal D}(G\cup G_f,M)$ 
also provides a very convenient framework 
above which one may study the equivalent classes of formal context.


\begin{thebibliography}{99}  
\bibitem{LLJD12-1}
T.M. Liaw, S.C. Lin, A General Theory of Concept Lattice (I): 
Emergence of General Concept Lattice, to submit.   
\bibitem{Wi82}
R. Wille, Restructuring lattice theory, In Rival, I., editor, Ordered Sets. Reidel, Dodrecht. (1982) 445–470.
\bibitem{GW99}
B. Ganter, R. Wille, Formal Concept Analysis: Mathematical Foundation, Springer (1999).
\bibitem{Wi05}
R. Wille, Formal Concept Analysis as Mathematical Theory
of Concepts and Concept Hierarchies in B. Ganter et al. (Eds.): Formal Concept Analysis, LNAI 3626 (2005) 1–33.
%\bibitem{Pa82}
%Z. Pawlak, Rough sets. International Journal of Computing and Information Sciences. 18 (1982) 341–356
%\bibitem{Pa91}
%Z. Pawlak, Rough sets, Theoretical Aspects of Reasoning About Data, Kluwer Academic Publisher 1991
\bibitem{Ke96} R.E. Kent, Rough concept analysis: a synthesis of rough sets and formal
concept analysis, Fundamenta Informaticae, 27, 169-181, 1996.
%\bibitem{DGO01}
%I. D{\" u}ntsch, G. Gediga, E. Orlowska, Relational attribute systems, International 
%Journal of Human-Computer Studies. 55 (2001) 293–309
\bibitem{GD02}
G. Gediga, I. D{\" u}ntsch, Modal-style operators in qualitative data analysis,
Proceedings of the 2002 IEEE International Conference on Data Mining, (2002) 155-162.
\bibitem{DG03}
I. D{\" u}ntsch, G. Gediga, Approximation operators in qualitative
data analysis, Theory and Application of Relational Structures as
Knowledge Instruments, de Swart, H., Orlowska, E., Schmidt, G. and
Roubens, M. (Eds.), Springer, Heidelberg, (2003) 216-233.
%Yao, Y.Y. A comparative study of formal concept analysis and rough set theory in
%data analysis, Rough Sets and Current Trends in Computing (RSCTC’04), LNAI
%3066, 59-68, 2004.
\bibitem{YY04}
Y.Y. Yao, Concept lattices in rough set theory, 
Processing NAFIPS '04, IEEE Annual Meeting of the Fuzzy Information, Vol.2 (2004) 796-801. 
%\bibitem{Wa05}
%P. Wasilewski, Concept Lattices vs. Approximation Spaces in
%D. Slezak et al. (Eds.): RSFDGrC (2005), LNAI 3641, 114–123
%\bibitem{BH00}
%S. N. Burris, H. P. Sankappanavar, A Course in Universal Algebra. Springer-Verlag, the Millennium Edition (1981)
%\bibitem{Bi67}
%G. Birkhoff, Lattice Theory, third edition, Amer. Math. Soc. Providence R.I. (1967)
\bibitem{Ku01} S.O. Kuznetsov, On Computing the Size of a Lattice and Related Decision Problems; Order 18, 4 (2001), 313-321.
\bibitem{KO02} S.O. Kuznetsov, S.A. Obiedkov, \q{Comparing 
performance of algorithms for generating concept lattices}; J. Exp. Theor. Artif. Intell. 14, 2-3 (2002), 189-216.
\bibitem{GD86}%
J. L. Guigues and V. D. Duquenne,
Familles minimales d'implications informatives r{\' e}sultant
d'un tableau de donn{\' e}es binaires
Math{\' e}matiques et sciences humaines, tome 95 (1986) p5-18.
\bibitem{Kso04}
S. O. Kuznetsov, On the intractability of computing the 
Duquenne-Guigues base, J. Universal Comput. Sci. 10 (8) (2004) 
927-933.
\bibitem{KO08}
S. O. Kuznetsov and S. A. Obiedkov, Some decision and counting problems
of the duquenne-guigues basis of implications, Discrete Applied Mathematics, 156(11), (2008) 1994-2003.
\bibitem{Sb09}%
B. Sertkaya, Towards the complexity of recognizing pseudo-intents, In F. Dau and
S. Rudolph, Eds, ICCS 2009, Lecture Notes in Computer Science, vol. 5662, Springer-Verlag, (2009) 284-292.
\bibitem{Df10}%
F. Distel, Hardness of enumerating pseudo-intents in the lectic order, in: L. Kwuida, B. Sertkaya (Eds.), ICFCA 2010, in Lecture Notes in Artificial
Intelligence, vol. 5986, Springer, (2010) 124-137.
\bibitem{DS11}
F. Distel, B. Sertkaya, On the complexity of enumerating pseudo-intents, Discrete Applied Mathematics 159 (6) (2011) 450-466.
\bibitem{BK13}
M. A. Babin, S. O. Kuznetsov,
Computing premises of a minimal cover of functional dependencies
is intractable, Discrete Applied Mathematics 161 (2013) 742-749.
%\bibitem{Aww74}
%W. W. Armstrong, Dependency Structures of Data Base Relationships, IFIP Congress (1974) 580-583
\bibitem{LLJD12-3}
T.M. Liaw, S.C. Lin,  
A General Theory of Concept Lattice (III):
From Categorisation to Logic Reasoning, to submit. 
\bibitem{Hd27}
D. Hilbert, \q{The foundations of mathematics}, translated by Stephan Bauer-Menglerberg and Dagfinn F{\o}llesdal (pp. 464-479) in 
\q{From Frege to G{\"o}del: A Source Book in Mathematical Logic}, 
1879-1931,
J. Van Heijenoort (ed.) Harvard University Press (1967).  
\bibitem{LLJD12-4}
T.M. Liaw, S.C. Lin, 
A General Theory of Concept Lattice (IV):
Resolvable Primary Logic Deduction System, to submit.
\bibitem{LLJD12-5}
T.M. Liaw, S.C. Lin, 
A General Theory of Concept Lattice (V):
Toward Practical Algebraically Deducible Logic Reasoning, to submit.
\end{thebibliography}
\end{document}